\documentclass[useAMS,aas_macros]{mn2e} 
\usepackage{aas_macros} 
\usepackage{amssymb}
 \usepackage{epsfig} 
 \usepackage{aastex_hack} 
 \usepackage{multirow} 

\newcommand{\etal}{et~al.~} 
\newcommand{\Sersic}{S\'{e}rsic }

\newcommand{\kms}{\ifmmode\,{\rm km}\,{\rm s}^{-1}\else km$\,$s$^{-1}$\fi} 
\newcommand{\magarc}{\ifmmode {{{{\rm mag}~{\rm arcsec}}^{-2}}} 
             \else {{{mag}$~${arcsec}$^{-2}$}} 
             \fi} 
                                                                                 
\def\Equnp#1{Eq.~\ref{eq:#1}} 
\def\Equ#1{Eq.~(\ref{eq:#1})} 
\def\se#1{\S\ref{sec:#1}} 
\def\Fig#1{Fig.~\ref{#1}}

\def\be{\begin{equation}} 
\def\ee{\end{equation}} 
\def\ifm#1{\relax\ifmmode#1\else$\mathsurround=0pt #1$\fi}

\def \spose#1{\hbox to 0pt{#1\hss}} 
\def \lta{\mathrel{\spose{\lower 3pt\hbox{$\sim$}} 
     \raise 2.0pt\hbox{$<$}}} 
\def \gta{\mathrel{\spose{\lower 3pt\hbox{$\sim$}} 
     \raise 2.0pt\hbox{$>$}}} 
\def\ltsima{$\; \buildrel < \over \sim \;$} 
\def\lsim{\lower.5ex\hbox{\ltsima}} 
\def\gtsima{$\; \buildrel > \over \sim \;$} 
\def\gsim{\lower.5ex\hbox{\gtsima}}

\def\kms{\ifmmode\,{\rm km}\,{\rm s}^{-1}\else km$\,$s$^{-1}$\fi}

\def\c28 {C$_{28}$}

\def \V22{V_{2.2}}

\def \ion#1#2{#1{\footnotesize{#2}}\relax} 
 
\def \hi {\ion{H}{I}\ }

\title[Structure of Virgo Cluster Galaxies]{The Near-IR Luminosity
  Function and Bimodal Surface Brightness Distributions of Virgo
  Cluster Galaxies}

\author[M. McDonald, S. Courteau \& R.~B. Tully]{Michael McDonald$^{1,*}$, 
 St\'{e}phane Courteau$^1$, \& R. Brent Tully$^2$\\
\\
$^1$Department of Physics, Engineering Physics and Astronomy, 
 Queen's University, Kingston, ON, Canada\\
$^*$Currently at University of Maryland, College Park, MD\\
$^2$Institute for Astronomy, University of Hawaii, 2680 Woodlawn Drive, Honolulu, HI\\
mcdonald@astro.umd.edu, courteau@astro.queensu.ca, tully@ifa.hawaii.edu} 

\begin{document}

\pagerange{\pageref{firstpage}--\pageref{lastpage}} \pubyear{2009}

\maketitle 
 
\label{firstpage} 

\begin{abstract}
We have acquired deep, H-band, imaging for a sample of 286 Virgo cluster 
galaxies with B$_T \leq 16$ mag and extracted surface photometry from 
optical $g$,$r$,$i$,$z$ Sloan Digital Sky Survey images of 742 Virgo 
Cluster Catalog galaxies, including those with H-band images. 
We confirm the detection of a dip in the luminosity function indicative 
of a discontinuity in the cluster galaxy population; the dip is more 
pronounced at redder wavelengths.  We find, in agreement with earlier
works of Tully \& Verheijen and ours for Ursa Major cluster galaxies, 
a clear dichotomy between high and low surface brightness galaxy disks. 
The difference between the low and high brightness peaks of Virgo disk 
galaxies is $\sim$2 H-mag arcsec$^{-2}$, significantly larger than any 
systematic errors. The high surface brightness disk galaxies have two 
distinct classes of high and low concentration bulges, while low 
surface brightness galaxies have only low concentration bulges.  
Early-type galaxies exhibit a similar structural bimodality though
offset from that of the spiral galaxies towards higher surface 
brightnesses.  Both the early- and late-type structural bimodalities 
are uncorrelated with colour or any other structural parameter except,
possibly, circular velocity.  Random realizations of realistic surface
brightness profiles suggest that a bimodal distribution of effective
surface brightness is unexpected based on normal distributions of
bulge and disk parameters.  Rather, the structural bimodality may be
linked to dynamical properties of galaxies. 
Low angular momentum systems may collapse to form dynamically
important disks with high surface brightness, while high angular
momentum systems would end up as low surface brightness galaxies
dominated by the dark halo at all radii.  The confirmation of
structural bimodality for gas-rich and gas-poor galaxies in the
high-density Virgo cluster as well as the low-density UMa cluster
suggests that this phenomenon is independent of environment.
\end{abstract}

\section{Introduction}\label{sec:intro}
The range of surface density profiles for galaxies in a cluster 
is a telltale of its dynamical history.  For instance, an excess 
of cuspy profiles over, say, the mean field galaxy distribution, 
is indicative of recent merger activity.  Conversely, extended 
exponential profiles are representative of quiescent evolution
over long timescales (e.g., Toth \& Ostriker 1992).  The study of 
galaxy light profiles has a rich history with seminal early 
contributions from de Vaucouleurs (1948; 1959), \Sersic (1968), 
and Freeman (1970).  De Vaucouleurs (1948) established that the 
surface brightness (hereafter SB) profiles of early type stellar 
systems may follow a strongly concentrated distribution now 
referred to as the ``de Vaucouleurs'' or ``$r^{1/4}$'' profile.  
De Vaucouleurs (1959) later reported the ubiquity of exponential 
SB profiles in disk galaxies while \Sersic (1968) proposed a 
generalized fitting function that encompasses the exponential and de
Vaucouleurs profiles. 

Based on photographic 
images for 36 disk and S0 galaxies, Freeman (1970) postulated 
that the distribution of disk central surface brightness (hereafter 
CSB), defined as the intercept of an exponential disk fit at $r=0$, 
peaks at $\mu_0$=21.65 B mag arcsec$^{-2}$.  Deep, wide-field CCD, 
galaxy surveys have since revealed a rather continuous distributions
of CSBs from high surface brightness (HSB) to low surface brightness
(LSB) galaxies (de Jong \& Lacey 2000)\footnote{Besides the small
number statistics, Freeman's study was biased by the low threshold 
of photographic plates to bright galaxies.}. Simulations and 
observations of galaxy structure have thus far suggested a continuous
range of properties over, rather than fundamental departures from, 
the HSB and LSB regimes. However, based on near-infrared (NIR) 
observations of the Ursa Major (UMa) cluster, Tully \& Verheijen 
(1997; hereafter TV97) inferred that the distribution of disk 
central surface brightnesses could be bimodal.

Based on a sample of 62 Ursa Major cluster galaxies observed at optical 
and NIR bands, TV97 argued that the distribution of disk central surface
brightnesses, $\mu_0$, was not continuous as expected from simple structure
formation models, but rather bimodal. The bimodality of the $\mu_0$
distribution was not convincingly observed at optical (BRI) bands, which 
TV97 argued might be due to extinction effects at optical bands. 
At K$^{\prime}$\footnote{The K$^{\prime}$ filter was described by
  Wainscoat \& Cowie (1992). It resembles the 2MASS Ks, with the
  primary difference being that it excludes the longest wavelength
  part of the K atmospheric window in order to reduce thermal
  emission.}, TV97 could assert the existence of two peaks in the SB
distribution at 17.28 K$^{\prime}$ $\magarc$ for HSB galaxies
(corresponding to ``Freeman's law'') and at 19.69 K$^{\prime}$
$\magarc$ in the LSB regime. The surface brightness of the gap was
centered at $\sim$18.5 K$^{\prime}$ $\magarc$.  Inclination and
extinction corrections to the optical surface brightness profiles
would bring a CSB bimodality to light in these bands as well. TV97
stressed, however, that the SB bimodality at optical bands could
plausibly be an artifact of their extinction correction which applies
only to HSB galaxies.  Still, the fact that the K$^{\prime}$-band is
nearly insensitive to dust and that the distribution of
$\mu_{0,K^{\prime}}$ values is clearly bimodal is a strong case for
TV97's argument of structural bimodality in spiral galaxies.

TV97 divided their sample into galaxies with and without a significant
near-neighbour, to identify an environmental effect.  The small UMa 
membership however means that the isolated and crowded sub-samples 
contain only roughly 30 galaxies each.  Despite the statistical
limitations, TV97 reported evidence for an enhancement of the 
bimodality in the isolated sub-sample.  The majority of galaxies 
with intermediate $\mu_0$ measurements were found to have near 
neighbours. This led TV97 to suggest the existence of two stable
dynamical configurations of baryonic and dark matter, likely
induced by the environment, leading to the distinct LSB and 
HSB populations. 

Comparing light distributions with total mass distributions
inferred from \hi synthesis maps allowed TV97 to further classify 
these two states.  In LSB galaxies, dark matter dominates the 
potential at all radii; gas and stars have sufficient angular 
momentum that they have not settled to the core and are orbiting 
in response to the dark matter distribution.  In HSB galaxies, 
baryonic matter has dissipated its energy and transferred angular 
momentum sufficiently to become self-gravitating over the central 
$\sim{2}$ exponential disk scale lengths (Courteau \& Rix 1999; 
Dutton \etal 2007, hereafter D07).  A gap between LSB and HSB 
galaxies may suggest that galaxies avoid a situation where 
baryonic and dark matter have comparable dynamical influence 
in the inner disk.  The higher incidence of intermediate surface 
brightness (hereafter ISB) galaxies with near neighbors suggests 
that tidal influences might drive a galaxy from the LSB to the 
HSB state.

Although TV97's study provided several valid arguments for a
structural bimodality of cluster galaxies, it still suffered 
a few shortcomings. First, the UMa cluster membership is  
small.  With only 62 galaxies, a structural study is plagued 
by statistical uncertainty (Bell \& de Blok 2000). Second, TV97 
estimated the disk central surface brightness, $\mu_0$, by 
fitting an exponential profile to the outer disk in order to 
avoid the bulge component.  The uncertainty in this 
``marking-the-disk'' technique depends on the subjective 
interpretation of the bulge size and the disk fit baseline.  
This caveat was however examined carefully by McDonald \etal 
(2008) and found not to be the cause of the observed bimodality.  
Finally, TV97's result could be unique to the UMa cluster and 
not representative of the true nature of disk galaxies
in clusters or the field (de Jong \& Lacey 2000).  Because the 
UMa cluster lacks early type galaxies, the observed bimodality 
could simply be due to a missing morphological class of galaxies. 
It is, however, our impression that none of these objections are 
stronger or more convincing than the claim of bimodality itself, 
as we shall verify below for Virgo cluster galaxies. 

Following the release of the Sloan Digital Sky Survey 
(York \etal 2000; hereafter SDSS), there have been many 
reports of bimodality in the distributions of galaxy color, 
star formation and clustering properties (Strateva \etal 2001; 
Blanton \etal 2003; Kauffmann \etal 2003; Baldry \etal 2004; 
Brinchmann \etal 2004; Balogh \etal 2004) and even reports 
of trimodality in galaxy concentrations (Bailin \& Harris 2008). 
Colors, star formation and clustering of galaxies are all 
intimately linked with a transition baryonic (light-weighted) 
mass of 3$\times$10$^{10}$M$_{\odot}$ (Kauffmann \etal 2003). 
Red galaxies, at the upper end of the mass scale typically have 
low current star formation rates and are found primarily in 
cluster or high-density environments, while the less massive blue 
galaxies have high star formation rates and are found primarily in 
the field (Dressler 1980). These results have catalyzed several new
variants of galaxy formation models and evolution.  For instance, 
an appealing explanation for any star formation bimodality
involves two phases of gas accretion in galaxies via hot and cold
modes (Birnboim \& Dekel 2003; Dekel \& Birnboim 2006). The hot mode 
is the standard picture of gas that is shock heated to the virial
temperature during gravitational collapse of the gas and halo. This
shock-heated gas eventually cools and settles into a galaxy (Rees \&
Ostriker 1977; White \& Rees 1978; White \& Frenk 1991). The cold mode
consists of cold filaments penetrating far inside the halo (Birnboim
\& Dekel 2003). These can occur in low-mass halos where virial shocks
are absent and cold gas is accreted quasi-spherically. For halos with
larger masses, the gas is shock heated to the virial temperature
everywhere except where cold, dense gas filaments penetrate the virial
radius. As the mass increases, only filaments of increasingly dense
cold gas can penetrate the halo without being shock-heated. Kere{\v s}
\etal (2005) used semi-analytic models to predict that the transition
{\em halo} mass where hot accretion becomes more efficient than cold
accretion is $\sim$$3\times10^{11}$M$_{\odot}$. This concords with the
transition {\em baryonic} mass, $3\times10^{10}$~M$_{\odot}$, in the
color/SFR bimodality, assuming a universal baryonic to dark matter
ratio of $\sim$10\% (Zavala \etal 2003). Kere{\v s} \etal (2005) 
also found that low-mass halos in dense environments have enhanced hot
accretion. The cold mode would thus apply mostly to late-type, disk galaxies
while the hot mode pertains mostly to early-type, spheroidal, galaxies. 
This simplistic picture appears to match observations well, but 
realistic star formation prescriptions are needed for more 
robust data-model comparisons.  

\subsection{Plan of action}

We now wish to test TV97's claim of structural bimodality in 
a different environment and using a larger sample.  Thanks to 
its proximity, large size, and broad morphological coverage, 
the Virgo cluster is the next logical cosmic structure to
consider as we revisit TV97's analysis.  If confirmed, we 
can also assess whether the reported bimodality of surface 
brightnesses, for spiral galaxies at least, is in any way 
related to the SDSS color bimodality discussed above and 
whether the global trimodality of galaxy concentrations for
all Hubble types (Bailin \& Harris 2008) is also observed. 


The organisation of this paper is as follows: In \se{sample}, we
discuss briefly our database for the 742 Virgo Cluster Catalog
(Binggeli \etal 1985; hereafter VCC) that are found in the SDSS 6th
Data Release (Adelman-McCarthy \etal 2008; hereafter DR6).  Out of
these, we have defined a complete, magnitude-limited, sample of 286
VCC galaxies with $B_T \leq 16$ for which deep H-band photometry was
obtained. The full description of our data set, and details about the
data acquisition and reduction methods are given in McDonald \etal
(2009; hereafter ``data paper'').  We introduce in \se{analysis}
specific parametric and non-parametric quantities that will permit a
comprehensive understanding of the structure of Virgo cluster galaxies
of all types.  In Appendix A, we determine the effects of various
errors on the distributions of various non-parametric quantities in
\se{models}.  We present final results in \se{results}, and examine
their possible interpretations in \se{discussion}.

We assume in this paper a distance to all Virgo cluster galaxies 
of 16.5 Mpc or m-M=31.18 (Mei \etal 2007).  At that distance, $1\arcsec$ = 80 pc.  

\section{Virgo Sample}\label{sec:sample}

\begin{figure}
\centering
\includegraphics[width=0.48\textwidth]{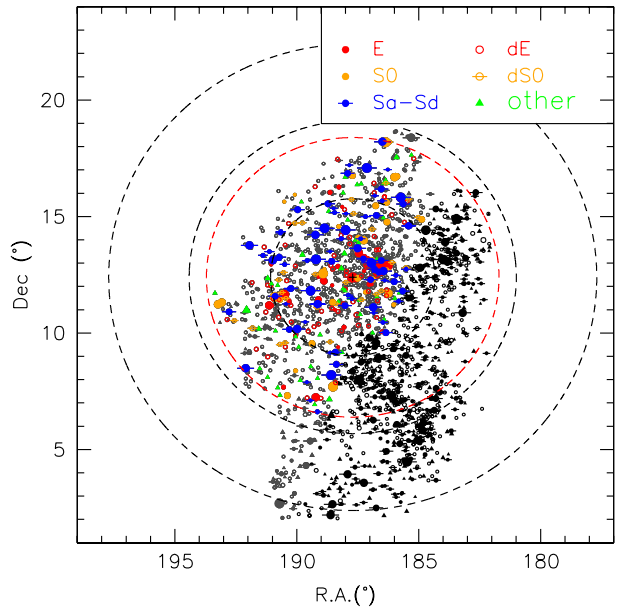}
\caption{Two-dimensional map of the Virgo cluster of galaxies. 
Colored points are galaxies that belong to our H-band sample
and that lie at a distance of $\sim 16.5$ Mpc. Grey points are
galaxies in our VCC/SDSS sample, while black points are the remaining
galaxies in the VCC. The size of the points scales with the total luminosity. The concentric dashed circles are projected
galactocentric distances (1, 2 and 3 Mpc) around the center of the
Virgo cluster at M87. The dashed red circle corresponds to a
distance of $6^\circ$ from M87.}
\label{virgomap}
\end{figure} 

\begin{figure*}
\centering
\includegraphics[width=0.7\textwidth,angle=270]{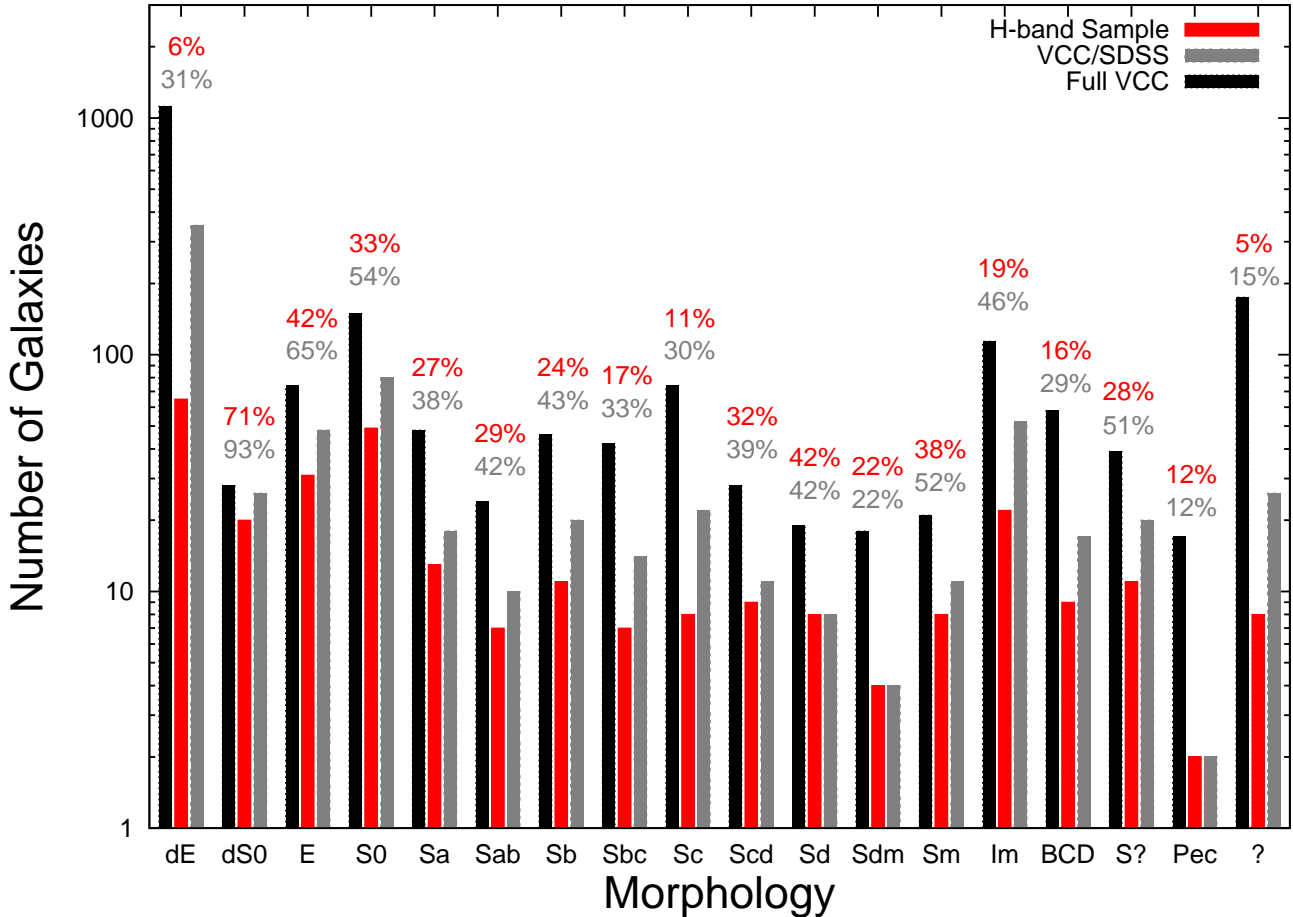}
\caption{Distribution of galaxy morphologies in the full VCC catalog
 (black), in the VCC/SDSS sample (grey), and in the H-band sub-sample
 (red).  The percent ratios are the number of objects in a given 
 morphological bin for each of the two samples divided by that 
 in the full VCC.}
\label{virgo_morph_hist}
\end{figure*}

The study of the unbiased distribution of galaxy surface brightness
requires volume completeness. A solution to this issue is to study
galaxies at a common distance in a cluster.  Our sample is drawn 
from the Virgo Cluster catalogue (Binggeli \etal 1985; hereafter VCC).  
The VCC catalogue contains 2096 galaxies within an area of 
$\sim$140 deg$^2$ on the sky, centered on the galaxy M87 at 
$\alpha$$\sim$12$^h$25$^m$ and $\delta$$\sim$13$^{\circ}$. 
The VCC is asserted to be complete down to a limiting absolute 
magnitude of $M_B \sim -13$ and to contain many objects as faint 
as $M_B \sim -11$ (Binggeli \etal 1985). 

Our first task was to acquire deep, optical, photometry for as many
VCC galaxies as could be found in the SDSS/DR6.  A further spatial cut
was made following Trentham \& Tully (2002) to reject 40\% of the
cluster that is contaminated by the W, W' and M background groups. The
W, W' and M groups were identified by de Vaucouleurs (1961) and
Ftaclas et al. (1984), respectively.  W and M lie about twice as far
away as Virgo and W' lies about 50\% further than Virgo.  These and
related structures in the Virgo Southern Extension all lie in a
flattened plane close to the supergalactic equator, contaminating the
western edge of the Virgo Cluster.  Trentham \& Tully (2002) discuss
this problem and describe how the projected area of the cluster can be
separated between zones with negligible and serious contamination.

Images from the SDSS/DR6 were extracted for a total of 742 VCC 
galaxies obeying these spatial cuts.  We will refer to this sub-sample 
as the ``VCC/SDSS'' sample. 

In order to ensure that our Virgo catalog reaches below the LSB peak
of UMa (TV97), we must achieve completeness at least down to
$M_B=-16.45$ mag (assuming a distance of 15.5 Mpc to UMa).  In order
to ensure that no intermediate surface brightness galaxies are missed,
we can define a complete sub-sample that includes all VCC galaxies
with $M_B \geq -15.15$ mag (i.e. $B_T \leq 16$). We also make an
additional two spatial cuts to ensure that we only include bound
cluster members: 1) all galaxies identified as ``background'' by
Binggeli \etal (1985) are removed and 2) all galaxies further than
$6^\circ$ from the center of the cluster, defined as the position of
M87, are removed.  These magnitude and spatial cuts leave us with a
total of 303 VCC galaxies centered on M87.  A further 8 VCC galaxies
(1068,1217,1258,1355,1665,1768,1889 and 2096) with recessional
velocity measurements, $V_{rad}>$ 3000 \kms, and the galaxies VCC0723
and VCC0991, with significant foreground stars, were also excluded.
The following seven VCC galaxies (530,950,1052,1287,1571,1822 and
1992) could also not be detected at H-band (see \se{nir}), and were left 
out of the sample (this has no effect on our final conclusions).  
We are left with a final, complete, sub-sample of 286 VCC galaxies
with $B_T \leq 16$ , covering a wide range of luminosities and 
morphologies.  For reasons that will soon be clear, we will
refer to this complete sub-sample as our ``H-band'' sample. 

\Fig{virgomap} shows the distribution of all VCC galaxies 
in black, as well as the 742 VCC/SDSS galaxies 
in grey, and the H-band sample in multi-colors. 
\Fig{virgo_morph_hist} shows the distribution of galaxy 
morphologies, as taken from the NASA Extragalactic Database 
(hereafter NED).  The broad morphological coverage is important 
to ensure that no distribution of galaxy structural parameter 
is biased by morphological segregation. 
\Fig{virgo_morph_hist} shows that Virgo is intrinsically rich
in dE, S0 and Im galaxies (see also Mei et al. 2007).  This 
is not due to a bias in our sample but simply to the nature of 
the Virgo cluster (as represented by the VCC).  

The gas-rich VCC galaxies are represented at approximately the same
levels in our VCC/SDSS and H-band samples relative to the full VCC
with completeness between 11\% and 42\%.  The bright magnitude limit of the 
H-band sample also implies a higher number of early-type 
galaxies relative to the later types.  However, we will 
show in \se{results} that the bimodality of surface brightnesses
is detected in each morphological bin (early or late type) and
that it is thus unlikely due to the morphological make-up of 
this specific cluster or our sampling of it. 

Following TV97's study of the UMa cluster, we have sought to 
obtain optical and especially near-IR photometry as well as 
dynamical measurements for all of our H-band sample galaxies. 
The SDSS $ugriz$ imaging for this sample will yield a distribution 
of luminosities, surface brightnesses, scale lengths and 
concentrations, as a function of optical wavelength. 
The optical colors will enable a comparison of any surface
brightness bimodality, if present, with the observed SDSS 
galaxy color bimodality (e.g., Strateva \etal 2001) and the ability 
to determine if one is simply a consequence of the other. 
The near-IR images are however essential to uncover the true 
distribution of galaxy surface brightnesses, luminosities, 
and other galaxy parameters largely free of extinction by dust.
We describe the extraction of the optical and NIR imaging data below.

While most critical, our collection of dynamical parameters for 
VCC galaxies is still in progress and will be reported elsewhere. 

\subsection{SDSS Photometry}\label{sec:SDSS}

We have extracted calibrated $ugriz$ images from the 
SDSS/DR6 for 742 VCC galaxies, including the 286 galaxies 
in our ``H-band'' sample.  Surface brightness profiles 
and total luminosities were obtained for these galaxies 
in all five SDSS bands by, first, performing isophotal 
ellipse fitting to the $i$-band images according to the 
methods of Courteau (1996), and then, applying the $i$-band 
isophotal solutions to the images in the other SDSS bands. 
The latter ensures that color gradients extracted from all 
SDSS images are computed from the same matching isophotes 
(MacArthur \etal 2003).  $u$-band images were consistently 
shallower than the $griz$ bands and were thus discarded. 
Sky levels for background subtraction and the photometric 
zero-points for calibration were obtained from the SDSS 
image headers and the SDSS archives, respectively. 
The remainder of the profile extraction technique is 
identical to that used for the near-IR photometry, 
as described below. 

\subsection{NIR Image Collection}\label{sec:nir}

Due to practical constraints, new, deep, H-band imaging that 
would sample well below the putative surface brightness 
bimodality scale of TV97 could only be obtained for a 
smaller sample of VCC galaxies.  This is the 
magnitude-limited ``H-band'' sample of 286 VCC galaxies
with SDSS imaging described above. 

Deep H-band imaging for some VCC galaxies is already 
available from the Two Micron All-Sky Survey 
(Skrutskie \etal 2006; hereafter 
2MASS\footnote{\tt{http://www.ipac.caltech.edu/2mass/releases/allsky/}})
and from the GOLDMine\footnote{\tt{http://goldmine.mib.infn.it/}} 
database (Gavazzi \etal 2003).
We were able to secure H-band imaging for the remainder of
the H-band sample with the detectors ULBCAM at the UH~2.2-m 
telescope, WFCAM at UKIRT and WIRCAM at CFHT over the period 
2005-2008.  Some deep K-band imaging was also available for
a few galaxies in 2MASS and GOLDMine. 

H-band images from GOLDMine were kindly provided by G. Gavazzi.
Calibrated 2MASS galaxy images were extracted from the online
database.  Many of the 2MASS and GOLDMine images were not deep enough
for our purposes.  Whilst adequate for large HSB galaxies, the high
2MASS brightness threshold (typically $\mu_H = 21$ mag
arcsec$^{-2}$. Bell \etal 2003; Courteau \etal 2007; Kirby \etal 2008)
limits the use of those data bases for deep extragalactic
studies. Likewise, just a handful of GOLDMine images were deep enough
to properly separate the bulge and disk light.  We have defined the
relative depth criterion, $Q$, as the ratio of the maximum extent of
the H-band surface brightness profile divided by that of the SDSS
$i$-band profile: $Q=r_{\rm{max,NIR}}/r_{\rm{max},i}$, where $r_{max}$
is the radius where the surface brightness error exceeds 0.15
$\magarc$ (see McDonald \etal 2008 for details).  Wherever possible,
we impose $Q>0.75$ for our NIR data.
Ultimately, 20 2MASS and 79 GOLDMine galaxy profiles were 
deemed useable for our study.  A final 
187 Virgo cluster galaxies needed new observations. These new and
existing observations are summarized in Table \ref{obstable}.

\begin{table}
\caption{Summary of H-band observations for the 286 Virgo cluster galaxies in our sample.}
\centering
\begin{tabular} {cccc}
\hline\hline
Tel - Camera & Targets & Collected & Avg Seeing\\
\hline
UH88$"$ - ULBCAM & 52 & 04/2005 & 1.2 $\pm$ 0.2 \\
UH88$"$ - ULBCAM & 16 & 04/2006 & 1.5 $\pm$ 0.3 \\
UH88$"$ - ULBCAM & 31 & 04/2007 & 1.3 $\pm$ 0.2 \\
UH88$"$ - ULBCAM & 23 & 03/2008 & 1.0 $\pm$ 0.2 \\\\
UKIRT - WFCAM & 31 & 07/2008 & 1.1 $\pm$ 0.1 \\
CFHT - WIRCAM & 34 & 02-06/2008 & 1.1 $\pm$ 0.1 \\
GOLDMine & 79 & - & 2.2 $\pm$ 0.9 \\
2MASS & 20 & - & 2.6 $\pm$ 0.1\\

\hline
\end{tabular}
\label{obstable}
\end{table}

The surface brightness profiles for the deep GOLDMine and 2MASS 
images were measured using the same techniques as our new NIR
images to ensure uniformity for the entire database. Further details
regarding the data reduction process and quality are presented in
McDonald et al. (2008; 2009).


\section{Surface Brightness Profile Analysis}\label{sec:analysis}

Surface brightness profiles were extracted for the 742 VCC/SDSS
galaxies at $griz$ bands and for the 286 H-band galaxies. In this
section, we examine the various parametric and non-parametric
properties derives from these profiles. 

McDonald \etal (2008) already considered the parametric (i.e. 
model dependent) analysis of UMa spiral galaxies in order to 
test, and ultimately confirm, TV97's claim of bimodality. 
The current, larger, Virgo cluster sample however includes galaxies 
of all morphologies.  While a parametric approach to model the shape of 
complex light profiles involves a multi-component decomposition 
(e.g., MacArthur \etal 2003; McDonald \etal 2008), a non-parametric 
approach is free of model assumptions and reveals different aspects
of galaxy structure.  We explore both approaches below and it will 
be shown later that surface brightness bimodality is essentially
independent of the method of light profile analysis. 

\subsection{Parametric Quantities}
Spheroidal and flattened galaxy systems have traditionally been 
modeled as the sum of a bulge and disk components (see 
MacArthur \etal 2003; McDonald \etal 2008 for more details).  
The 1D light profile of a galaxy disk is typically parametrized 
as an exponential function: 
\be
I_d(r)=I_0\exp\left\{-{r\over{h}}\right\},
\ee
where $I_0$ and $h$ are the disk central surface brightness and scale
length, respectively.  Meanwhile, the projected 1D bulge light profile 
is better modeled as a \Sersic function (\Sersic 1968):
\be
I_b(r)=I_e\exp\left\{-b_n\left[\left({r\over{r_e}}\right)^{1/n}-1\right]\right\},
\label{eq:sersic}
\ee
where $r_e$ is the half-light radius, $I_e$ is the surface brightness
at that radius, and $n$ is the \Sersic shape parameter.
With $n=1$, the \Sersic function reduces to the exponential function.



In addition to the bulge and disk, we consider other components
that may affect the light profile such as compact nuclei, spiral arms 
and disk truncations.  Ignoring these may result in large errors
in the bulge and disk parameters (McDonald \etal 2008).


B/D decompositions were performed only for the H-band and optical $griz$ 
light profiles of the 286 galaxies in the ``H-band'' sample (McDonald \etal 2009).

\subsection{Non-Parametric Quantities}

Non-parametric quantities are measured directly from the surface 
brightness profile, with no prejudice for any assumed model. 
The only assumptions inherent to the non-parametric measurements below 
are: (i) that the total galaxy light is an extrapolation of the light
profile to infinity, and (ii) that inclination estimates and extinction
corrections are valid.  Non-parametric quantities allow a direct, 
unbiased comparison of galaxies across the full Hubble sequence.

\subsubsection{Concentration, $C_{28}$}
The galaxy light concentration is a measure of the relative 
light fraction between the inner and outer parts of the galaxy.  
Unlike the B/D ratio, which relies on a model for the light distribution, 
$C_{28}$ is a straightforward, model-independent morphological indicator.
The concentration, $C_{28}$ is defined as (e.g., Kent 1985; 
Courteau 1996): 

\be
C_{28} \equiv 5\log\left({r_{80}\over{r_{20}}}\right)
\label{eq:c28}
\ee
where $r_{80}$ and $r_{20}$ are the radii enclosing 80 and 20
percent of the total light.  For a pure exponential function ($n=1$ in
\Equ{sersic}), $C_{28}$=2.8.

Concentration indices are a function of wavelength and while 
fractional radii depend on inclination and extinction corrections, 
we expect the ratio $r_{80}$/$r_{20}$ to be roughly independent 
of projection effects.  We justify this assumption below.

\subsubsection{Fractional Radii and Surface Brightness}
Fractional parameters refer to specific quantities measured 
at radii that contain specific fractions of the total galaxy 
light. The effective radius, $r_e$, is the radius that 
encloses half of the total light.  While $r_e$ was already
introduced in the context of a parametric profile function 
(\Equnp{sersic}), it is formally defined in a non-parametric 
way as:
\be
\int^\infty_0 I(r)2\pi rdr \equiv 2\int^{r_e}_0 I(r)2\pi rdr.
\label{eq:reff}
\ee
The effective surface brightness, $I_e$, is the surface brightness 
at $r_e$; $I_e=I(r_e)$ ($\mu_e \equiv -2.5$log$I_e$).  Unlike the
extrapolated central surface brightness of the disk, $I_0$, the 
effective surface brightness $I_e$ is non-parametric, making it 
ideal for the comparison of mean surface brightness levels for galaxies 
of varying morphology; $I_e$ (or $\mu_e$) applies to all galaxy types 
rather than $I_0$ (or $\mu_0$) which is restricted to spiral disks
($I_0=1.678I_e$ for pure disks).  
Our discussion about the variations in galaxy surface brightness 
profiles will indeed rely on $r_e$ and $\mu_e$.

Another fiducial light marker is the fractional radius defined as:

\be
\int^{r_x}_0 I(r)2\pi rdr \equiv {x\over{100}}\int^\infty_0 I(r)2\pi rdr
\ee

where $r_x$ contains some percentage ($x$) of the total light. 
$\mu_x$ is the surface brightness in $\magarc$ at $r_x$.  A final,
non-parametric, measure of surface brightness is the average surface 
brightness interior to some radius $r_x$, $\langle I_x \rangle$, defined as:

\be
\langle I \rangle_x \equiv {\int^{r_x}_0 I(r)2\pi rdr\over{\pi r_x^2}}. 
\ee

$\langle \mu \rangle_x$ is the magnitude equivalent of $\langle I \rangle_x$. 

\smallskip

\begin{figure}
\centering
\includegraphics[width=0.48\textwidth] {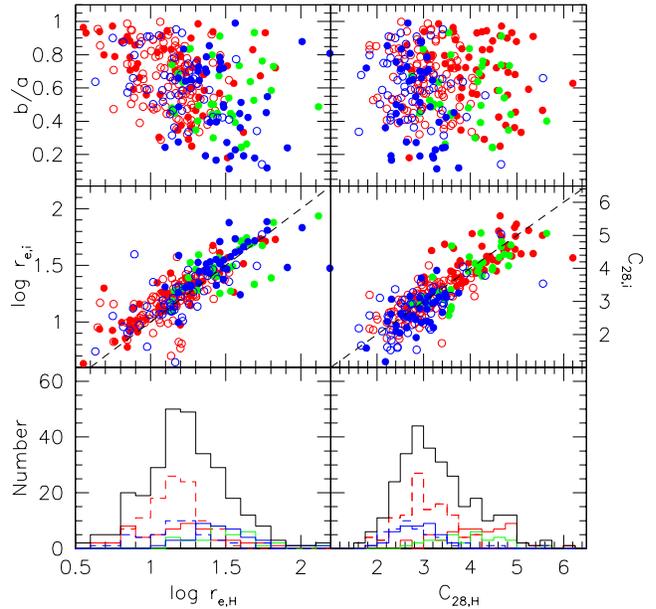}
\caption{Distribution of H-band effective radius, $r_{e,H}$, and
  concentration, $C_{28}$ for the Virgo H-band sample as a function of
  $i$-band axial ratio (top), and against similar $i$-band measurements
  in the middle section. The point/line types represent: dE-dS (red, 
  open/dashed), E-S0 (red, closed/solid), Sa-Sb (green, closed/solid),
  Sc-Sd (blue, closed/solid) and Irr (blue, open/dashed). The black, 
  dotted, line in the middle windows is the slope unity correlation. 
  The bottom windows are population histograms.}
\label{re_C_iH}
\end{figure}

\Fig{re_C_iH} shows the dependence of the H-band $r_{e,H}$ 
and $C_{28}$ on the $i$-band axial ratio, $b/a$ for the 
H-band sample.  There appears to be a correlation between 
$r_{e,H}$ and $b/a$, but it is driven largely by the 
spheroidal nature (high axial ratios) of compact galaxies
(small $r_e$).  Any correlation between $r_e$ and $b/a$ is 
significantly weakened if compact galaxies are excluded.  
The insensitivity of $C_{28}$ to projection effects (top
right window) is even clearer.  The respective corrections 
for projection on $r_{20}$ and $r_{80}$ roughly cancel out. 
We also see in \Fig{re_C_iH} that there is a good linear 
correlation between $i$- and H-band scale radii and 
concentrations. 

Ultimately, concentrations, fractional radii and surface brightnesses
were computed for all 742 VCC/SDSS at $gri$ wavelengths and for all
286 ``H-band'' galaxies at H-band.

\section{Results}\label{sec:results}

We now examine the distribution of structural parameters of Virgo cluster
galaxies, starting with an analysis of the gas-rich systems in the H-band 
sample for comparison with TV97's similar analysis.  We then derive the 
distribution of effective surface brightnesses and the optical and NIR 
luminosity functions for the full H-band and VCC/SDSS samples, with
all morphologies considered.

We will find for the Virgo gas-rich galaxies in the H-band sample,
a bimodal distribution of the extrapolated disk central surface 
brightness, $\mu_{0,H}$, and thus bolster the similar claim for
UMa galaxies by TV97.  For the complete sample of gas-rich and gas-poor
Virgo galaxies, we determine the distribution of effective surface 
brightness, $\mu_e$ (independent of B/D decompositions) and find
compelling evidence for brightness bimodality in each morphological
group.   We present in \se{csb} the distributions of $\mu_0$ and, 
in \se{esb}, various fractional and average surface brightness measures. 
We compute in \se{lumfuncs} the optical and NIR luminosity 
functions and compare these to the recently published optical 
luminosity function of Virgo (Rines and Geller, 2008) and 
field NIR luminosity function from UKIDSS (Smith et al. 2008). 
In \se{conc}, we present the distributions of various scale 
radii and the galaxy concentration, $C_{28}$. The latter is compared
to the distribution of concentrations for SDSS galaxies by Bailin and
Harris (2008). 
In \se{bivariate} we examine the bivariate distributions
for most of the structural parameters addressed in this section.

\subsection{Disk Central Surface Brightness}\label{sec:csb}

In order to compare with the study of UMa cluster galaxies by TV97 and
McDonald \etal 2008, we restrict our H-band sample to disk
galaxies. B/D decompositions, as in McDonald \etal (2008), were
performed for this sub-sample of 166 VCC disk galaxies.  The B/D fits,
which include optional bulge, nucleus and spiral arm components, were
applied to each $griz$ and H-band SB profiles.  The derived $\mu_0$
values were corrected for projection effects to their face-on value in
the absence of extinction. Thus, we write: $\mu_0^i\equiv\mu_0 -
2.5C_\lambda\log(b/a)$, where $b/a$ is the measured axial ratio of the
outermost isophote at a given band in each galaxy.  We take
$C_\lambda$=1 for the dust transparent case.  This is likely a fair
assumption at H-band, the focus of our analysis, but less adequate for
the $griz$ bands. We keep our analysis free of dust correction for
now.

\begin{figure}
\centering
\includegraphics[width=0.48\textwidth]{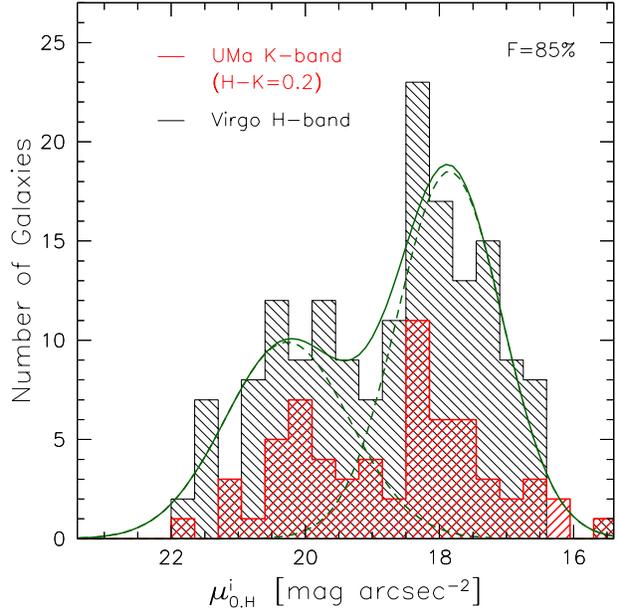}
\caption{Distribution of inclination-corrected disk central surface 
 brightnesses, $\mu_{0,H}^i$ for 166 VCC spiral and irregular galaxies (hatched). 
 The red line shows the results from the analyses of UMa galaxies by TV97 and
 ourselves (McDonald \etal 2008), while the green line shows the two-Gaussian 
 fit to the VCC data. The F-test value demonstrates that a single-Gaussian 
 fit can be rejected with a confidence of 85\%. The transition between the 
 two brightness peaks is at $\mu_{0,H}^i\sim 19\magarc$.}
\label{virgo_csb}
\end{figure}

\begin{figure}
\centering
\includegraphics[width=0.48\textwidth]{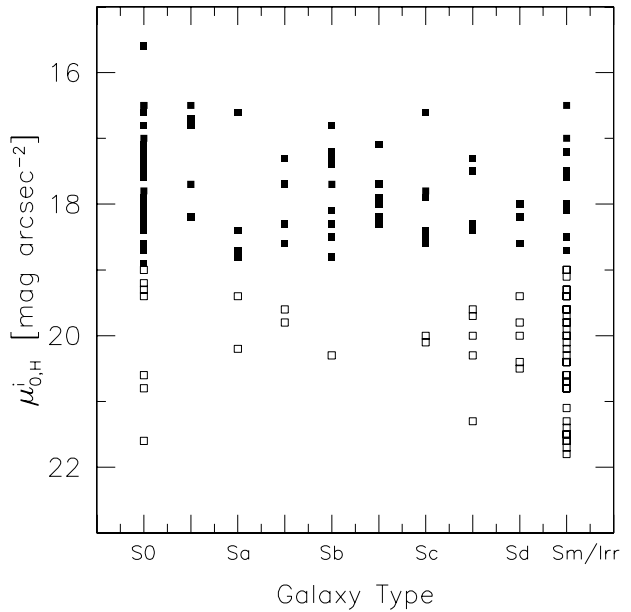}
\caption{Inclination-corrected disk central 
 surface brightnesses, $\mu_{0,H}^i$, with morphological type for 166
 VCC galaxies. Filled and open black squares represent HSB and LSB galaxies, 
 respectively.}
\label{csb-morph}
\end{figure}

The distribution of inferred $\mu_0^i$ values is shown in \Fig{virgo_csb}. 
There is a clear dearth of galaxies at $\mu_{0,H}^i\sim$19 mag arcsec$^{-2}$. 
This result for later-type VCC galaxies matches very well that determined 
by TV97 and ourselves for UMa disk galaxies with a minimum in the number of 
galaxies at $\mu_{0,K}^i\sim18.5$ mag arcsec$^{-2}$. Similarly, we find 
excellent agreement with TV97 in the location of the HSB and LSB peaks 
at $\mu_{0,H}^i$=17.85$\pm$0.15 mag arcsec$^{-2}$ and
$\mu_{0,H}^i$=20.27$\pm$0.4 mag arcsec$^{-2}$, respectively, 
with a peak separation of $\Delta\mu_{0,H}^i$=2.4 mag arcsec$^{-2}$.
Considering a typical H-K$\sim$0.2 for galaxy disks, the distributions 
for $\mu_{0,H}^i$ in the Virgo and UMa clusters are very similar.

Using a statistical F-test, we can reject the hypothesis 
of a normal distribution for $\mu_{0,H}^i$ in favor of a bimodal 
distribution with a confidence level of 85\%. \Fig{csb-morph} shows 
which galaxies contribute mostly to different surface 
brightness levels. As expected, early-type disk galaxies dominate the 
HSB peak, while late-type disk galaxies and irregulars are present 
in both the HSB and LSB peaks.

\begin{figure}
\centering
\includegraphics[width=0.48\textwidth]{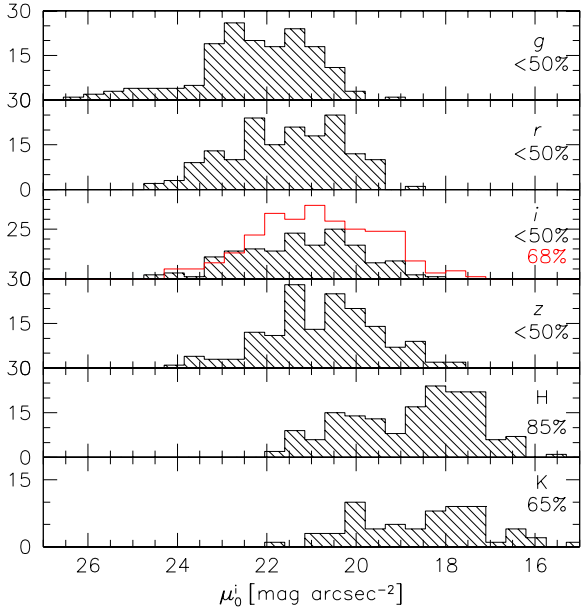}
\caption{Distribution of inclination-corrected disk central surface
  brightnesses, $\mu_0^i$, from optical (SDSS) to near-IR bands for
  166 spiral and irregular Virgo cluster galaxies from the H-band 
  sample. The red histogram in the $i$-band panel refers to the 272 disk
  galaxies in the VCC/SDSS sample. The numbers on the right are the
  statistical F-test probability for a 2-component Gaussian. }
\label{virgo_sdss_csb}
\end{figure}

Let us now examine the distribution of inclination-corrected $\mu_0$
with wavelength.  \Fig{virgo_sdss_csb} shows the different $\mu_0^i$
distributions from $g$ (top) to K (bottom).  The gap between the two
SB peaks grows as a function of wavelength since the HSB peak
brightens at a faster rate with wavelength than the LSB peak.  This is
explained by the higher dust content in HSB galaxies, in contrast with
their relatively transparent ($C^\lambda=1$) LSB counterparts (TV97),
as well as the relative colors of the constituent stars (HSB galaxies
tend to be redder than LSB galaxies).
Dust obscuration is less effective at longer wavelengths and, as 
a result, we see more deeply along any given line-of-sight thus 
increasing the observed surface brightness.  Recall that we have 
not applied any correction for obscuration (C$_{\lambda}$=1) to 
surface brightnesses and the bimodality disappears as we 
consider shorter wavelengths. It is possible that the correct 
choice of C$_\lambda$ would restore the bimodality at optical 
wavelengths, as suggested by TV97, but any such complications 
can be avoided by restricting our analysis to NIR wavelengths only, as
we do here.

\begin{figure}
\centering
\includegraphics[width=0.48\textwidth]{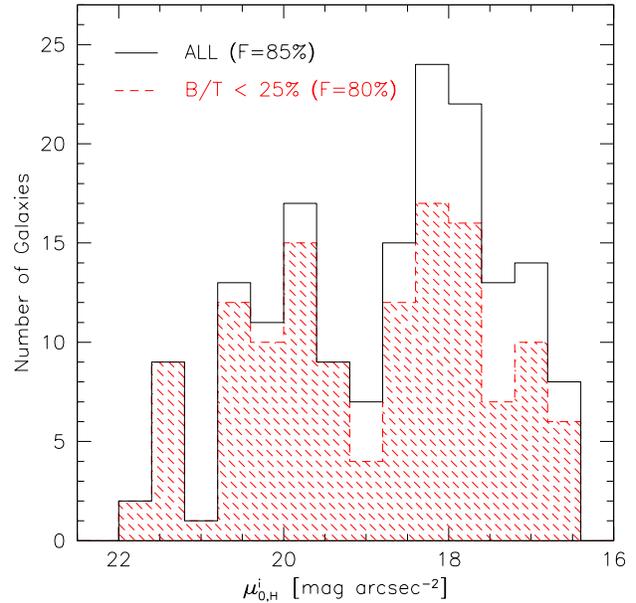}
\caption{Histogram of $\mu_{0,H}^i$ for 166 spiral and irregular VCC
  galaxies. The red, hatched histogram includes only those galaxies
  with very little contribution from the bulge component (bulge-to-total ratio, B/T, $<$ 25\%). This shows that bimodality is not the result of bulge-dominated vs bulgeless systems.}
\label{mu0_bulge}
\end{figure}

\begin{figure}
\centering
\includegraphics[width=0.48\textwidth]{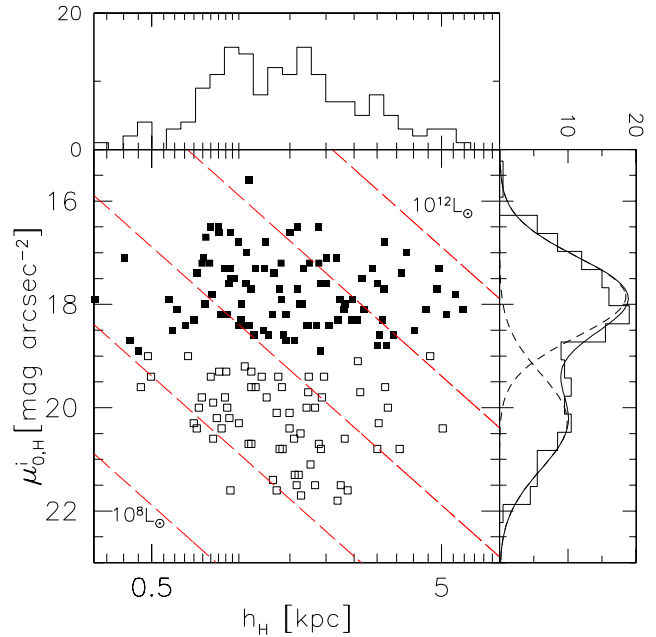}
\caption{Distribution of $\mu_0$ as a function of disk scale 
length, $h$, for the sub-sample of 166 spiral and irregular VCC galaxies. 
Red dashed lines represent constant luminosity; our detection
limit is $\sim$10$^9$L$_\odot$ at H-band. Filled and empty squares 
correspond to HSB and LSB galaxies respectively, as in \Fig{csb-morph}.}
\label{csb_h}
\end{figure}

\begin{figure}
\centering
\includegraphics[width=0.48\textwidth]{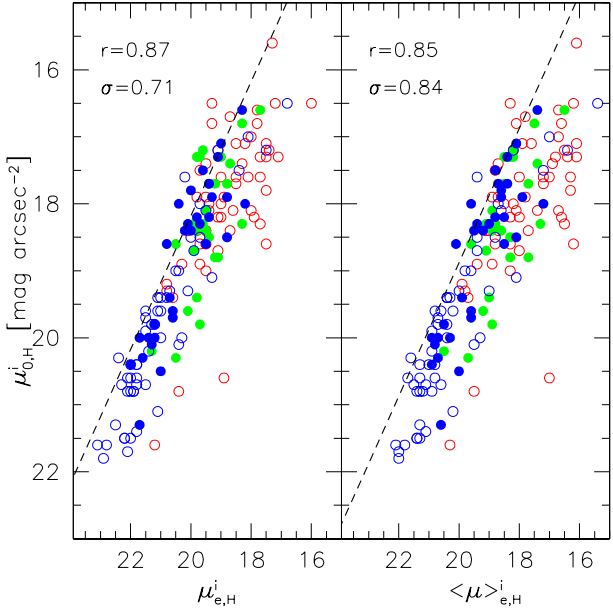}
\caption{Correlation between the parametric brightness, $\mu_0$, and
  non-parametric brightnesses $\mu_e$ and $\left\langle \mu\right \rangle_e$ for the
  166 gas-rich VCC galaxies. The
  point types are as follows: open red circle - S0, green circles - Sa
  \& Sb, blue circles - Sc \& Sd, open blue circles - irregulars. The
  dashed lines are $\mu_0=\mu_e-1.822$ and $\mu_0=\left\langle \mu\right \rangle_e-1.124$, 
  representing a pure exponential disk.}
\label{mu0-mue}
\end{figure}

We can ask if the bimodality in the disk CSB is
correlated with any bulge structure. Fig. \ref{mu0_bulge} shows the
same distribution as in Fig. \ref{virgo_csb} but now with a
second $\mu_{0,H}^i$ distribution including only galaxies with small
bulges. We define a ``small'' bulge as one which contributes
$<25$\% of the total galaxy light.  Indeed, we see that if we remove
those galaxies with bulges which contribute more than 25\% of the
total luminosity, bimodality is preserved. Though the F-test
confidence level drops, this is primarily due to the reduction in
sample size - the ISB gap is very well preserved.

\Fig{csb_h} shows the relationship between $\mu_0$ and disk scale
length, $h$, at H-band. At intermediate luminosity, there can be both HSB and LSB
galaxies, depending on the disk surface density (faint/high $\mu_0$ \& large
$h$ or bright/low $\mu_0$ \& short $h$). The fact that two galaxies with the
same luminosity can have wildly different surface brightnesses leads
to the belief that some mechanism likely related to the initial halo
angular momentum (Dalcanton \etal 1997) prevents LSB systems from 
collapsing to the same densities as the HSBs.
This figure makes clear that for a given luminosity the LSB galaxy
must be more extended than the HSB galaxy and that no LSB galaxy
exceeds 10$^{11}$L$_\odot$. The distribution of disk scale lengths
also shows some evidence of multiple peaks, though the statistical
significance is low (68\%), none of which are correlated with the
structure in $\mu_{0,H}^i$.

We have shown thus far that the NIR central surface brightnesses,
$\mu_{0,H}^i$ of disk galaxies in the Virgo and UMa clusters are distributed
bimodally.  Our B/D decompositions confirm a result that is already
well-known - that galaxy bulges come in two types: cuspy (high
concentration) and cored (low concentration). HSB disks harbour both
types of bulges, while LSB disks only harbour low-concentration
bulges.

We now expand our analysis to take full advantage of 
the diverse Virgo cluster population by considering early-type and 
dwarf galaxies as well.

\begin{figure}
\centering
\includegraphics[width=0.48\textwidth]{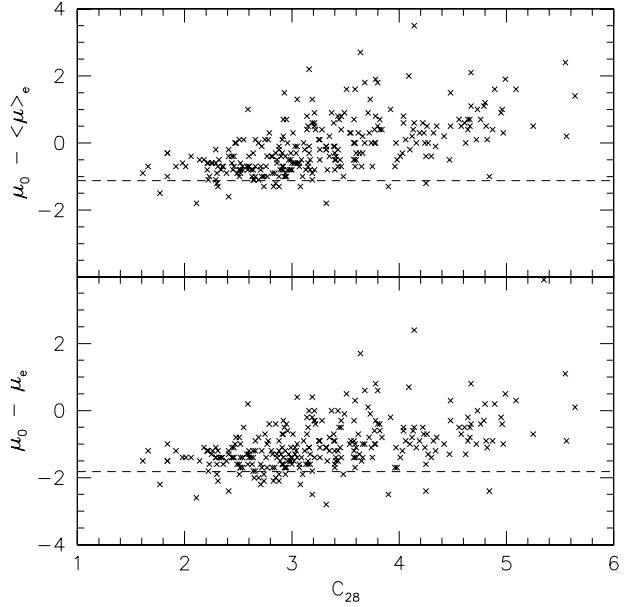}
\caption{$\mu_0-\mu_e$ and $\mu_0-\left\langle \mu\right \rangle_e$ residuals 
 as a function of concentration.}
\label{mu-resid}
\end{figure}

\subsection{Effective Surface Brightness}\label{sec:esb}
Unlike the UMa cluster of galaxies, Virgo is rich in giant and dwarf
early-type galaxies. The mean density of Virgo is $\sim$5 times that
of UMa.  Many of the VCC galaxies have no observable disk 
component and therefore our analysis of galaxy properties based 
on $\mu_0$ would be moot.  We consider instead two more versatile 
quantities: $\mu_e$, the galaxy effective surface brightness and 
$\left\langle \mu\right \rangle_e$, the mean surface brightness, defined as the
average surface brightness within $r_e$ (\S4.2.2).  These non-parametric 
quantities can be measured for all galaxies.  \Fig{mu0-mue} shows 
the comparison for the 166 gas-rich VCC galaxies of $\mu_0$ with $\mu_e$ 
and $\left\langle \mu\right \rangle_e$.  There is a strong correlation between 
$\mu_0$ and $\mu_e$ with the expected zero-point offset 
($\mu_e=\mu_0+1.822$) for pure exponential disks).  The scatter 
increases for earlier-type galaxies. This scatter is further examined in
Fig.\ref{mu-resid}, where we confirm that the differences between
$\mu_0$,$\mu_e$, and $\left\langle \mu\right \rangle_e$ are a function of
morphology or, for simplicity, concentration. That is, for higher
concentration the scatter in the $\mu_0-\mu_e$ and
$\mu_0-\left\langle \mu\right \rangle_e$ relations increases.

We have also considered the average surface brightness within $r_e$,
$\left\langle \mu\right \rangle_e$, as a measure of a galaxy's characteristic
surface brightness. Use of this parameter is motivated by a study of
late-type field galaxies by de Jong \& Lacey (2000; hereafter
DL00). DL00 studied the distribution of $\left\langle \mu\right \rangle_e$ for 1000
Sb-Sdm field and cluster galaxies which, they reported, is not
bimodal. The right side of \Fig{mu0-mue} shows that $\left\langle \mu\right \rangle_e$
correlates well with $\mu_0$, though with slightly larger
scatter than $\mu_e$. From this point on, we will adopt $\mu_e$ as our standard measure 
of surface brightness for three reasons: 1) For spiral galaxies, 
$\mu_e$ scales directly with $\mu_0$, 2) $\mu_e$ can be measured 
for any galaxy morphology, and 3) $\mu_e$ is independent of any
assumptions about the shapes of the galaxy bulges (if any) and disks. 
We also adopt a geometric correction for $\mu_e$ in all galaxies:
$\mu_e^i=\mu_e - 2.5\log(b/a)_{r_e}$, where $(b/a)_{r_e}$ is the axial
ratio at the effective radius. Regardless of what surface brightness
measure we use, and how we correct for projection effects, we will
show in \S6 that the overall shape of the distribution of surface
brightnesses is preserved.

\begin{figure}
\centering
\includegraphics[width=0.48\textwidth]{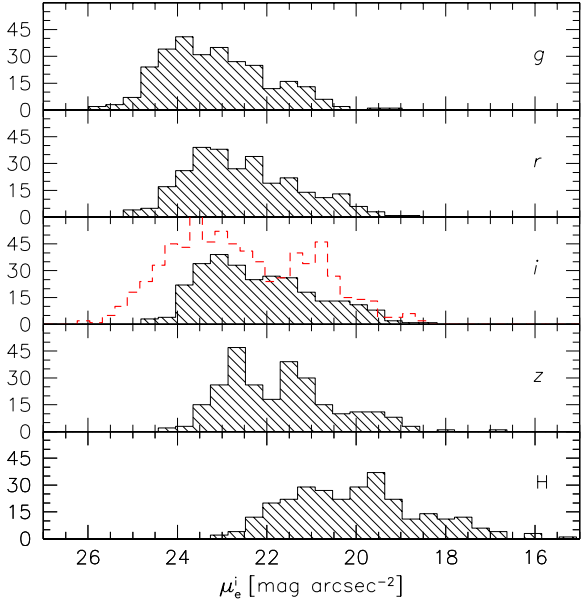}
\caption{Distribution of effective surface brightness, $\mu_e^i$, in 5
  bands for all 286 VCC galaxies in our sample. The red histogram in
  the $i$-band window is the distribution of $\mu_e^i$ for all
  742 VCC/SDSS galaxies; that distribution is strongly bimodal with an
  F-test of 95\%. Troughs are significant in the brightness
  distributions from $i$ to $H$ bands.}
\label{virgo_esb_sdss}
\end{figure}

We show in \Fig{virgo_esb_sdss} the distribution of $\mu_e^i$ 
with wavelength for all 286 H-band VCC galaxies. The differential
effects of extinction with wavelength are well known (though poorly
understood) and likely the cause for the smoother contributions at
shorter wavelengths. The $\mu_e$ distribution of the nearly dust-free
LSB galaxies should not change drastically from $g$ to H, modulo a
color term due to stellar populations. The HSB peak from $g$ to H
will, however, be altered by effective dust obscuration.  The faint
tail of the HSB peak will thus be stretched to fainter values,
effectively filling any intrinsic trough between the HSB and LSB
peaks (as shown by TV97 for UMa galaxies). For this reason, we shall
now rely solely on NIR surface brightness measurements for the
remainder of our analysis. 

\begin{figure}
\centering
\includegraphics[width=0.48\textwidth]{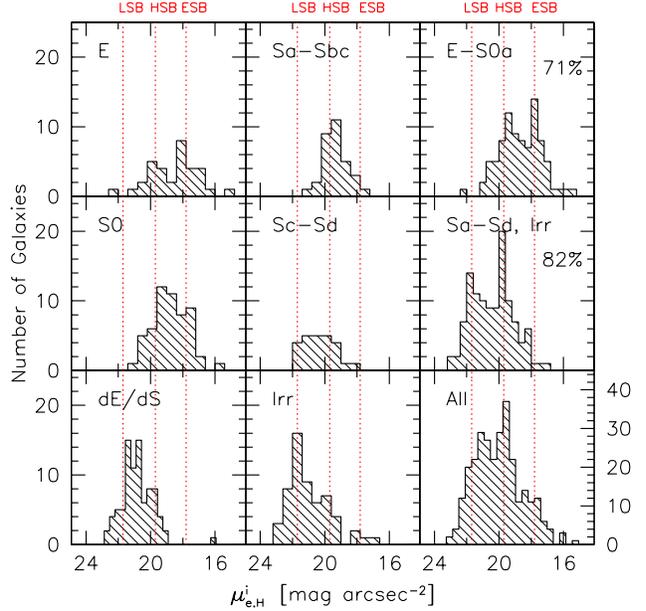}
\caption{Distribution of inclination corrected effective surface
  brightness, $\mu_e$, for 286 VCC galaxies of varying morphology.
  The red vertical dotted lines correspond to the locations of the ESB
  (extreme surface brightness), HSB and LSB peaks.  The upper-right
  panel shows all gas-poor galaxies, the middle-right panel is the sum
  of all the gas-rich galaxies, and the bottom-right is the sum of all
  morphological.  The sum of the two {\it bimodal} distributions for
  the gas-poor and gas-rich galaxy types respectively, results in a
  {\it trimodal} distribution of all the VCC galaxies in our
  sample. The numbers in the upper and middle right panels refer to
  the F-test confidence for bimodality.}
\label{virgo_esb_bymorph}
\end{figure}

\begin{figure}
\centering
\includegraphics[width=0.48\textwidth]{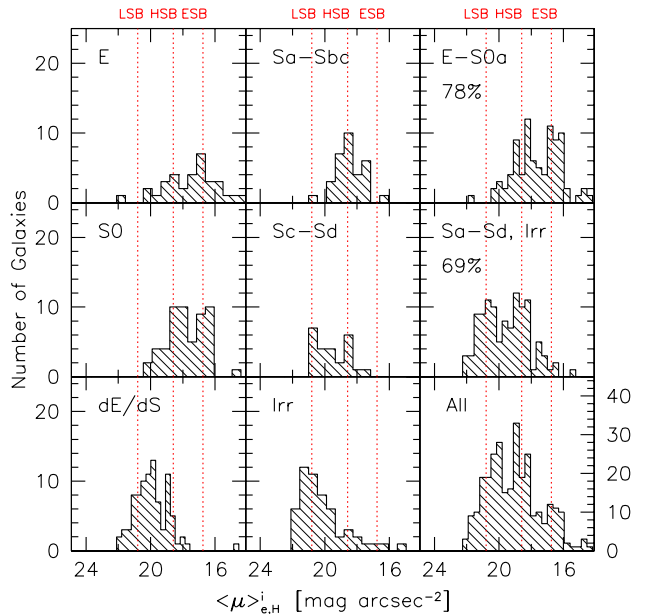}
\caption{Distribution of average surface brightness,
  $\left\langle \mu\right \rangle_e$, for galaxies of varying morphology. The numbers in
  the upper and middle right panels refer to the F-test confidence for
  bimodality.}
\label{virgo_avesb_bymorph}
\end{figure}

\Fig{virgo_esb_bymorph} shows the distribution of $\mu_{e,H}^i$
measurements for our sample of 286 H-band VCC galaxies.
Characteristic structure can be seen in most of the morphological
bins, as well as in the sample as a whole. The separate distributions
of E and S0 galaxies are each weakly bimodal. The combination of these
morphological types (top right panel of \Fig{virgo_esb_bymorph}) gives
a stronger bimodality for $\mu_{e,H}^i$. The bimodality of
$\mu_{e,H}^i$ for gas-rich galaxies emerges clearly from the HSB and
LSB peaks for the early-type (Sa-Sbc) and irregular galaxies; the
distribution of Sc-Sd galaxies shows the least features and is least
abundant of all VCC types. This results in a strong bimodality for all
the gas rich galaxies (middle-right panel of
\Fig{virgo_esb_bymorph}). If we compare the distributions of
$\mu_{e,H}^i$ for gas-poor and gas-rich galaxies, the lower surface
brightness spheroids line up with the HSB peak for late-type and dwarf
galaxies. This results in an \textit{apparent} trimodality of $\mu_e$
for the complete H-band sample (lower right panel of
\Fig{virgo_esb_bymorph}). However, we stress that the trimodality is
in fact a superposition of an early- and late-type SB bimodality. The
distribution of $\langle \mu \rangle_{e,H}^i$ in
\Fig{virgo_avesb_bymorph}, shows similar trends. However,
\Fig{virgo_avesb_bymorph} illustrates the sensitive nature of binning
for this sort of argument. For instance, the distribution of $\langle
\mu \rangle_{e,H}^i$ for Sc-Sd is here roughly bimodal when it was
flat in \Fig{virgo_esb_bymorph}. This illustrates the need for large
samples. We are still reassured that the distributions for $\mu_e$ or
$\langle \mu \rangle_e$ for the sum of both the gas-rich and gas-poor
galaxies show very similar (bimodal) features.
Figs.~\ref{virgo_esb_bymorph}-\ref{virgo_avesb_bymorph} show that a
structural bimodality is present in both early and late-type galaxies
with a tail towards the brightest galaxies. A statistical F-test
confirms the rejection of a single peak in favor of two distinct peaks
with 71\% and 82\% confidence for early- and late-type galaxies,
respectively. We can consider another significance test.  In order to
fill the early and late-type brightness gaps, an additional 15 and 29
intermediate SB galaxies would be needed, respectively.  However, the
current number of galaxies in these troughs is 24 and 39,
respectively, so the required additions to achieve flat SB
distributions would correspond to 3 and 4.5 standard deviations.  This
argument, based only on Poisson statistics, complements nicely the
F-test (which assumes a Gaussian shape).


Figs. \ref{virgo_esb_bymorph}-\ref{virgo_avesb_bymorph} tell an
exciting story; not only is a structural bimodality present in spiral
galaxies, but also in spheroids.  Equally intriguing is that spheroids
of lower surface brightness have identical surface brightnesses as 
HSB spirals, possibly pointing to a common formation and/or structural
mechanisms.  In order to better understand this result, we now examine 
the correlation of surface brightness with various other parameters 
and what role it plays in fundamental structural relations. But first, 
let us introduce another unexpected result, this time in the NIR luminosity
function, to shed more light on the structural bimodality in both
gas-poor and gas-rich galaxies.

\subsection{Near-IR and Optical Luminosity Functions}\label{sec:lumfuncs}
\begin{figure}
\centering
\includegraphics[width=0.48\textwidth]{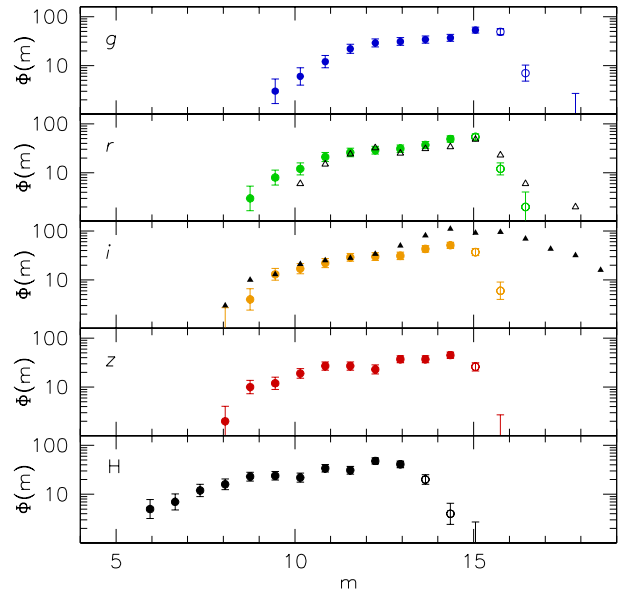}
\caption{Optical and NIR luminosity functions for our sample of 286
  Virgo cluster galaxies. In the $r$-band panel, the open triangles
  correspond to Petrosian magnitudes taken directly from the SDSS for
  comparison with Rines \& Geller (2008). In the $i$-band
  panel, the closed triangles correspond to magnitudes computed for
  all of the 742 VCC/SDSS galaxies. Open circles refer to magnitude bins
  suffering from incompleteness at the low-luminosity end. }
\label{lumfunc}
\end{figure}

McDonald \etal (2008) showed that the K-band luminosity
function for Ursa Major has a noticeable dip at intermediate
luminosities. We now examine this claim for Virgo galaxies with both
optical and near-IR luminosities. \Fig{lumfunc}
shows the luminosity functions of our 286 VCC galaxies in the
$g$,$r$,$i$,$z$ and H bands. There appears to be a dip in the
luminosity function at optical wavelengths, which develops into
a broad trough at H-band.  This trough clearly separates two 
distinct luminosity classes. 

\begin{figure*}
\centering
\includegraphics[width=0.9\textwidth]{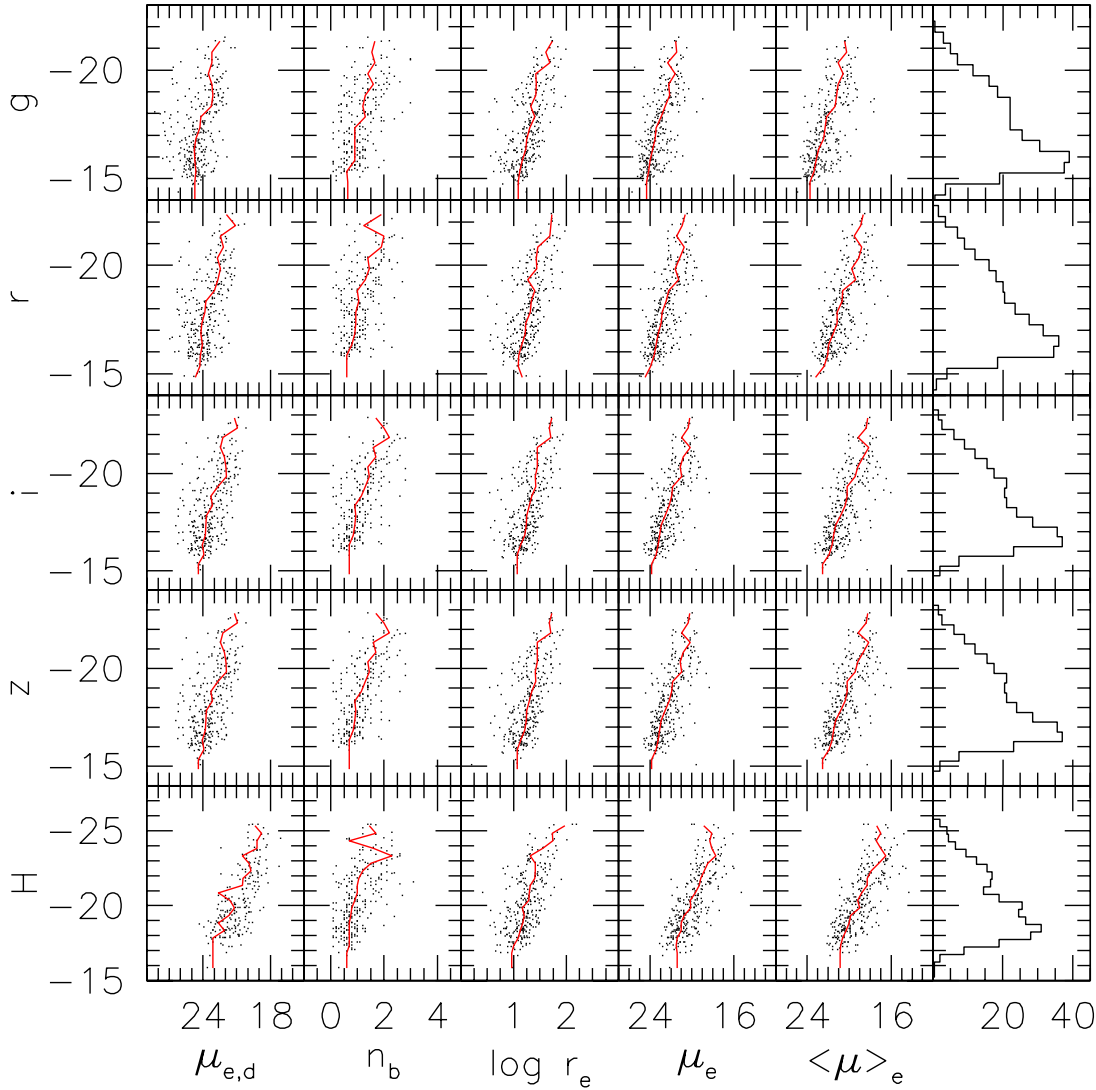}
\caption{Various structural parameters in X plotted against total
  absolute luminosity in Y for four optical bands ($griz$) and the
  H-band for 286 VCC galaxies. Structural parameters from left to
  right are as follows: $\mu_{e,d}$ - disk effective surface
  brightness, $n_b$ - Sersic index of the bulge, log $r_e$ -
  half-light radius, $\mu_e$ - effective surface brightness, $\langle
  \mu \rangle_e$ - average surface brightness. Histograms of the
  luminosity function are shown in the far right panels. The red lines
  represent the binned median using the same binning as the
  histogram.}
\label{everything}
\end{figure*}

Rines and Geller (2008) obtained the same result for their study of
the faint end of the Virgo LF (see black triangles in $r$-band window
of \Fig{lumfunc}), but dismissed it on grounds that SDSS data may
be flawed at luminosities where the dip is observed.  It would however
be challenging to think of such a systematic bias in the SDSS pipeline 
reductions.  Furthermore, our luminosities are measured from 
SDSS images but fully independent of any other SDSS data 
product (e.g., Petrosian luminosities and radii\footnote{Rines and
  Geller (2008) used ``model'' luminosities from the SDSS pipeline.}).  
The dip in the Virgo luminosity function is also observed 
at H-band which further confirms that it is not an artifact of the SDSS 
pipeline.  Our luminosity function and that of Rines \& Geller 
are a striking confirmation of each other, establishing the reality of
the dip in the luminosity function of Virgo cluster galaxies.

The dip in the NIR luminosity is not reproduced in the UKIDSS
luminosity function for field galaxies (Smith et al. 2008). This could
indeed mean that there is no trough in the luminosity function of
field galaxies and that the lack of intermediate-luminosity galaxies
in UMa and Virgo is purely a cluster-driven phenomenon. However, the
UKIDSS luminosity function suffers incompleteness at luminosities
fainter than $M_K=-20$, while the low-luminosity peak in the UMa
luminosity function is at $M_K=-19.5$. This fact, along with the
difficulty of achieving volume completeness in field galaxy samples,
could conspire to hide any putative trough in the NIR luminosity
function if it is, in fact, independent of environment.

We will see in \S4.5 and Fig. \ref{lum-mue} that the high-luminosity
peak is comprised of ``extreme'' surface brightness (hereafter ESB)
galaxies and the low-luminosity peak is comprised of LSB galaxies.
However, the HSB galaxies are distributed bimodally in luminosity,
implying a bimodality in sizes for these galaxies since they all have
nearly the same surface brightness.  The combination of Figs
\ref{virgo_esb_bymorph} and \ref{lumfunc} shows that the majority of
the low-luminosity HSB galaxies are compact dE galaxies, while the
majority of the high-luminosity HSB galaxies correspond to traditional
HSB spirals.

Finally, Fig. \ref{everything} shows the multi-dimensional variation
of various structural parameters with optical and NIR luminosity for
the full sample of 286 VCC galaxies. The dumbbell-shaped distribution
of points in all plots of luminosity versus $\mu_e$ or $\langle \mu
\rangle_e$ shows clearly the separate galaxy populations. The L-$\mu$
relation appears to be very similar whether we use $\mu_e$ or $\langle
\mu \rangle_e$, although the scatter is somewhat reduced at
$\mu_e$. Finally, there is a strong evolution of most of these scaling
relations with wavelength that will be studied elsewhere.

\subsection{Concentration}\label{sec:conc}
\begin{figure}
\centering
\includegraphics[width=0.48\textwidth]{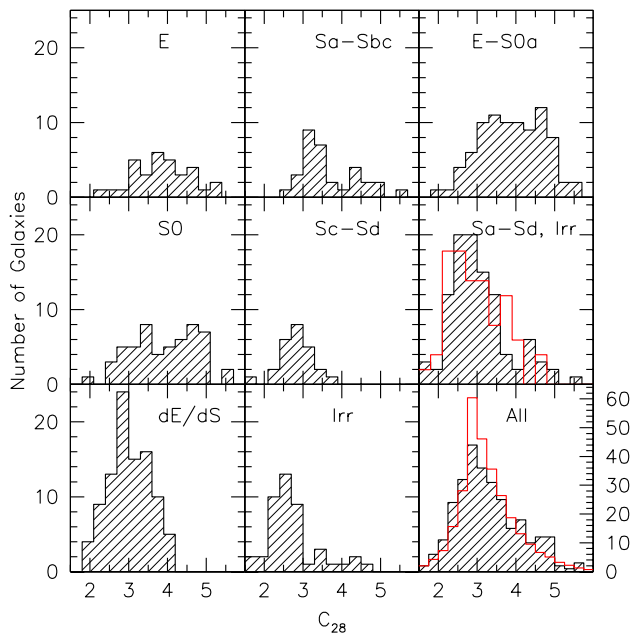}
\caption{Distribution of concentration, $C_{28}$, for 286 VCC galaxies
  of different morphologies. The red histogram in the Sa-Sd,Irr window
  refers to the distribution of $C_{28}$ for UMa galaxies from
  McDonald et al. (2008) for galaxies of the same morphology. The red
  histogram in the bottom-right window is based on 100,000
  model surface brightness profiles, discussed in \S5.4.}
\label{c28_bymorph}
\end{figure}

In Appendix A, we have attempted to predict the observed distribution of
$C_{28}$ based on normal distributions of $\mu_d$, $r_d$, $\mu_b$,
$r_b$ and $n$. In Fig. \ref{c28_bymorph} we show the distribution of
measured concentrations for the 286 VCC galaxies binned by morphology as well as
for the complete H-band sample. For comparison, we show the predicted
distribution from our models (Appendix A).

\begin{figure}
\centering
\includegraphics[width=0.48\textwidth]{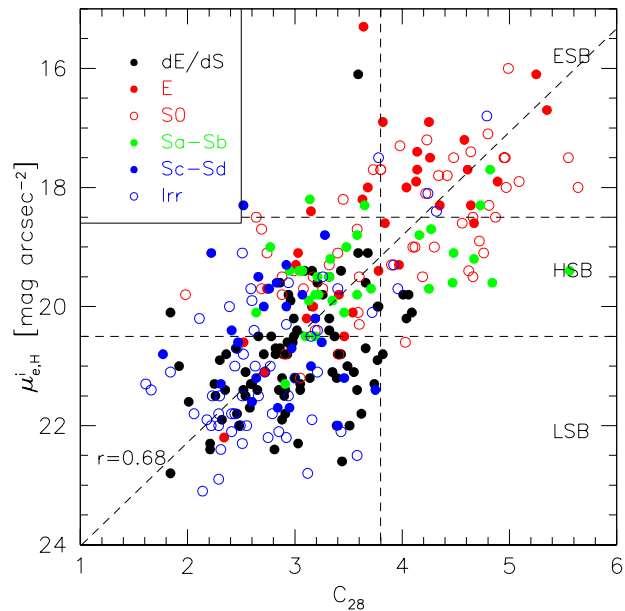}
\caption{Correlation between H-band concentration, $C_{28}$, and
  effective surface brightness, $\mu_{e,H}^i$. The dashed vertical
  line is an arbitrary transition at $C_{28}=3.8$ from low- to
  high-concentration, while the two dashed horizontal lines delineate
  the LSB, HSB and ESB classes.}
\label{c28_mue}
\end{figure}

The predicted and observed distributions are remarkably similar,
though the observation of a second peak at high concentration with
C$_{28}>4.0$ was not predicted by our simulations of isolated
galaxies. The overall distribution peaks at C$_{28}\sim3$,
corresponding to pure exponential disks. The distribution terminates
rather sharply at C$_{28}\sim2.0$, suggesting a lower limit for how
centrally concentrated the baryons in a galaxy can be. 
In order to understand the high-$C_{28}$ peak, let us look at 
the distribution of $C_{28}$ for different morphological types. 
\Fig{c28_bymorph} shows that the Sc-Sd, Irr and dE galaxies 
populate a single peak at $C_{28}<4$, while E, S0 and Sa-Sbc 
galaxies occupy two distinct peaks near $C_{28}=3$ and $C_{28}=4.5$.
Thus, there is a concentration 

bimodality only in early-type (bulge-dominated) galaxies. 
The difference between the model and observed distributions
is due in large part to the fact that our model distribution 
of bulge surface brightness is normal, while the one we observe 
is bimodal. The transition at $C_{28}=4$ represents the transition
from LSB to HSB bulges.  We combine Figs.~\ref{virgo_esb_bymorph} 
and \ref{c28_bymorph} in \Fig{c28_mue} to examine whether the
early-type bimodality in concentration corresponds to the 
bimodality in surface brightness. 

We see again in \Fig{c28_mue} that the two HSB peaks are comprised primarily of E, S0
and Sa galaxies, while the LSB peak is comprised primarily of dwarf,
irregular and late-type spiral galaxies. Most importantly, we see two
isolated clouds: one containing high-concentration HSB and ESB
galaxies and the other containing low-concentration HSB and LSB
galaxies. Again, LSB galaxies never have high concentration
and ESB galaxies never have low concentration. This provides some
insight into the structural bimodality in spheroids. It is possible
that galaxies with C$_{28}>4$ can only form through major mergers,
which naturally explains the lack of both LSB and star-forming galaxies 
in the high concentration peak. We can also verify that the
trimodality observed in SDSS concentrations by Bailin \& Harris (2008)
is not detected here.

\subsection{Bivariate Distributions}\label{sec:bivariate}
In order to assess if the brightness bimodality extends to, or even
stems from, other galaxy parameters, we now construct a number of
bivariate distributions for $\mu_e$ vs other fundamental structural
parameters such as color, luminosity and dynamical mass.

\begin{figure}
\centering
\includegraphics[width=0.48\textwidth]{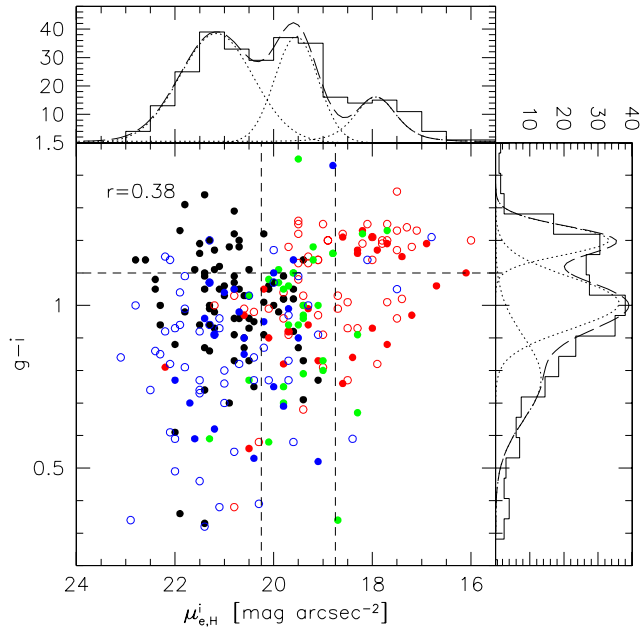}
\caption{Bivariate distribution of H-band surface brightness and $g-i$
  color for 286 VCC galaxies. Point colors and types are the
same as for Fig. \ref{c28_mue}. Dotted lines represent individual Gaussian fits while dashed lines are the sum of these fits.}
\label{mue_gi}
\end{figure}

\begin{figure}
\centering
\includegraphics[width=0.48\textwidth]{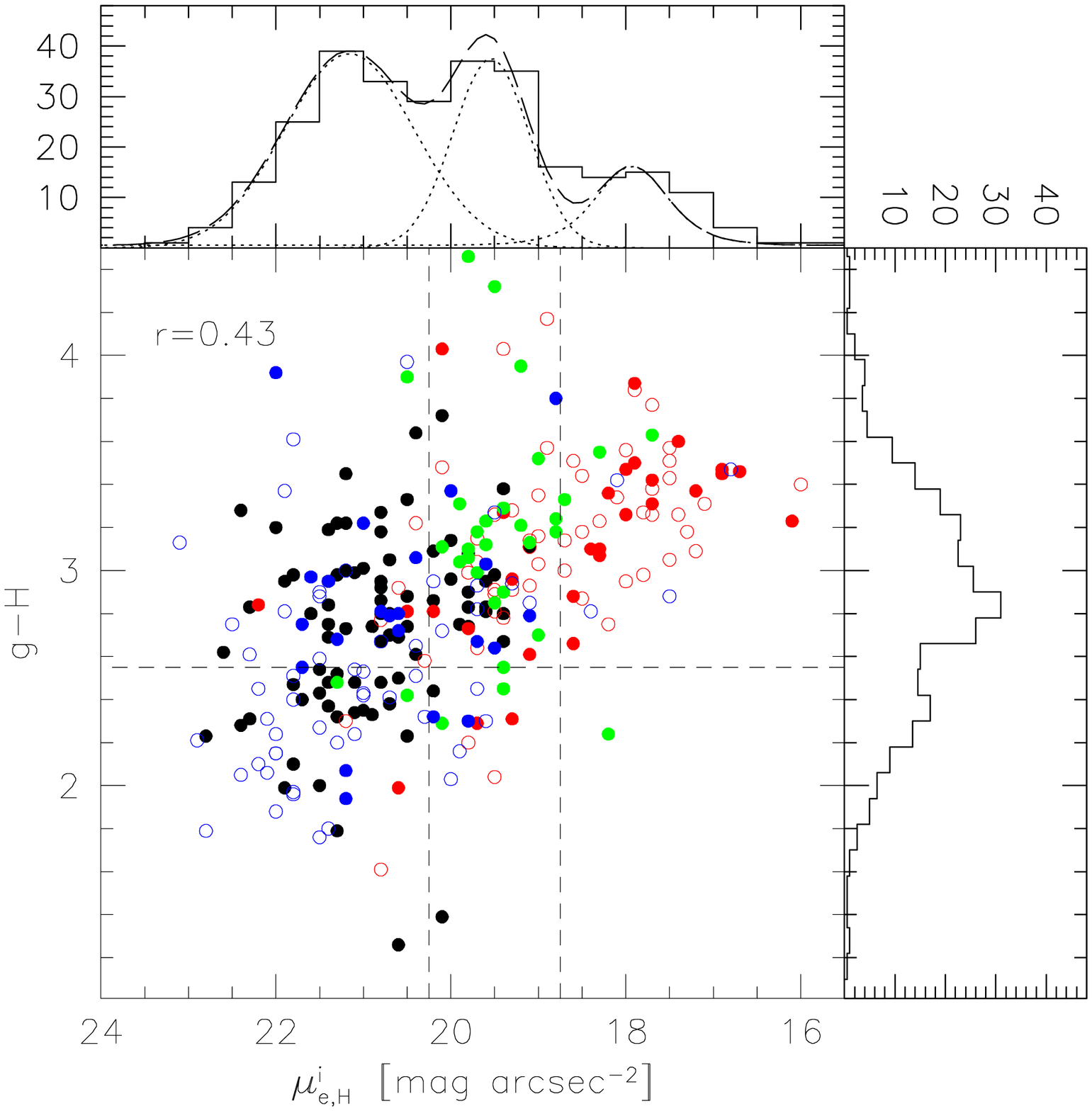}
\caption{Bivariate distribution of H-band surface brightness and $g$-H
  color for 286 VCC galaxies. Point colors and types are the
same as for Fig. \ref{c28_mue}. Dotted lines represent individual Gaussian fits while dashed lines are the sum of these fits.}
\label{mue_gH}
\end{figure}

\begin{figure}
\centering
\includegraphics[width=0.48\textwidth]{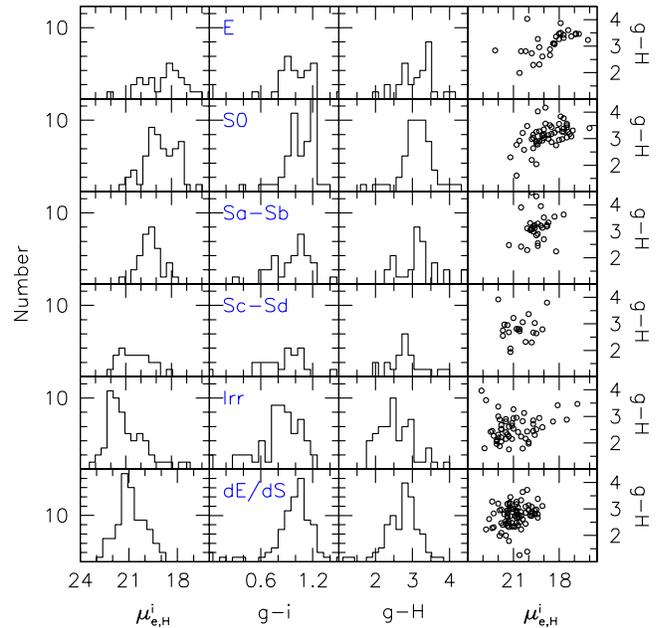}
\caption{Distribution of surface brightness and colors for
  286 VCC galaxies for different morphologies.}
\label{mue_col}
\end{figure}

The color-surface brightness bivariate distribution is certainly of
great interest in light of the color bimodality of SDSS galaxies
(e.g., Strateva \etal 2001). It is of obvious interest to ask if the
transition in galaxy colors matches the transition in galaxy surface
brightnesses. A careful examination of Fig \ref{mue_gi}
defuses this idea.
The fact that the blue and red peaks each
include all three SB peaks demonstrates that the structural
bimodality is indeed different from the color bimodality. The same
distribution but for $g$-H colors (Fig. \ref{mue_gH}) reveals a different color bimodality
(resulting from stellar population effects that will be investigated
elsewhere). The fact that the red peak alone is comprised of all three
SB classes again confirms that surface brightness trends and color
bimodality are independent phenomena. We summarize our findings in
Fig. \ref{mue_col} with the distributions of $g-i$, $g-$H and
$\mu_{e,H}^i$ for individual morphologies. There is a strong
correlation between color and surface brightness for elliptical
galaxies (and weaker for S0s), with redder ellipticals having higher
surface brightness than blue ellipticals. There appears to be no
correlation between color and surface brightness for the other types,
suggesting again that the surface brightness bimodality is independent
of stellar populations.

\begin{figure}
\centering
\includegraphics[width=0.48\textwidth]{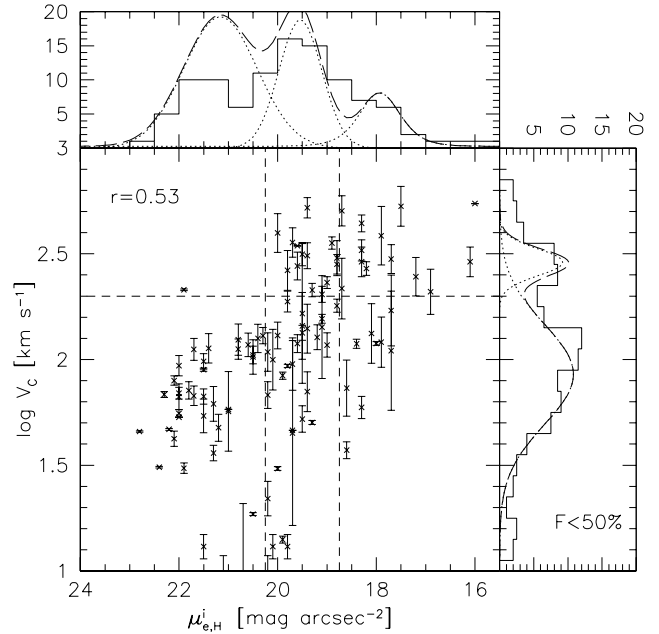}
\caption{Bivariate distribution of H-band surface brightness and
  fully corrected V$_{max}$ for 102 VCC spiral galaxies. The Gaussian fits in
  the upper panel correspond to the full distribution of 286 VCC galaxies. Dotted lines represent individual Gaussian fits while dashed lines are the sum of these fits.}
\label{vel-mue}
\end{figure}

\begin{figure}
\centering
\includegraphics[width=0.48\textwidth]{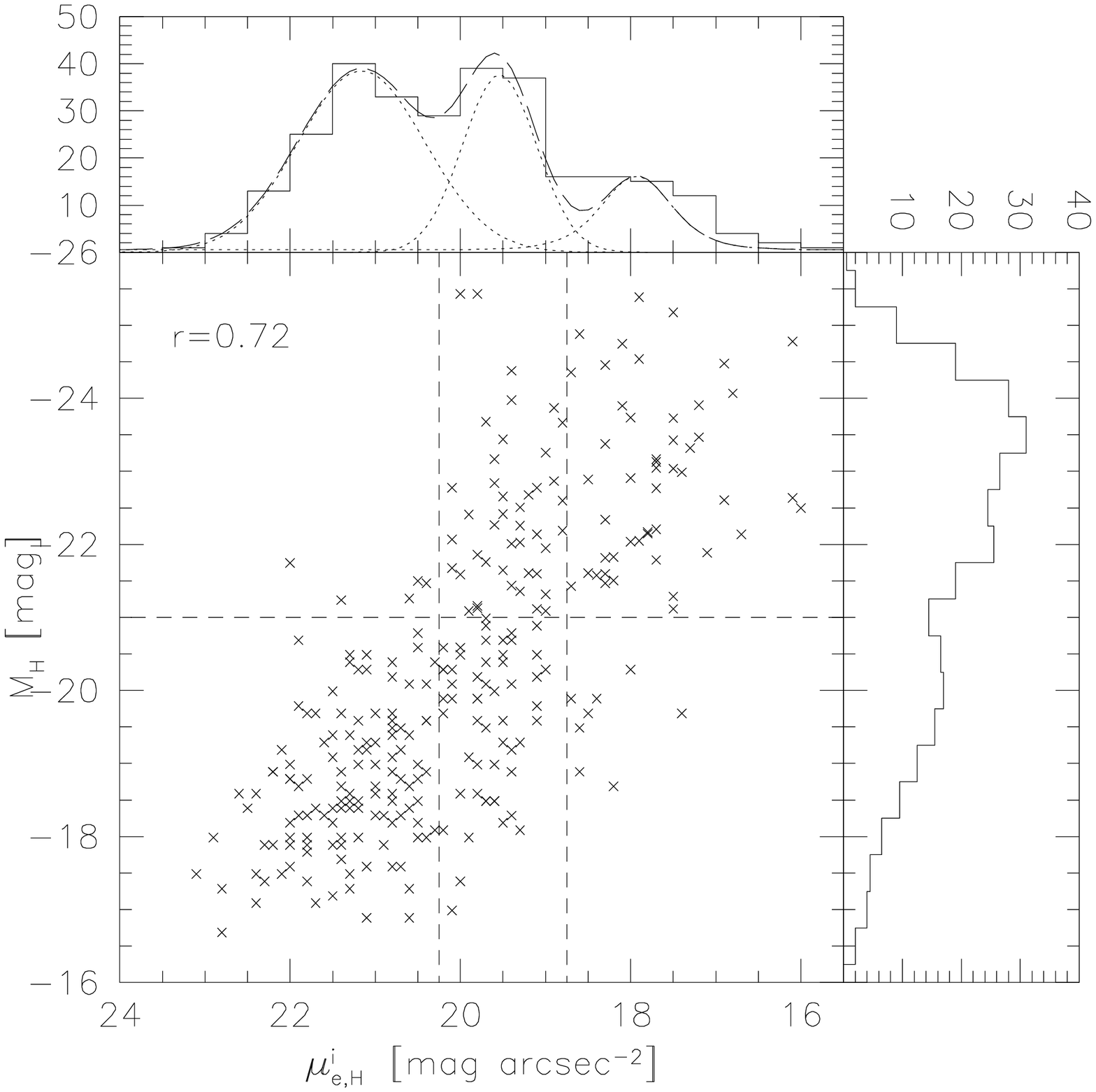}
\caption{Bivariate distribution of H-band surface brightness and total
  H-band luminosity for the 286 Virgo cluster galaxies with H-band
  photometry using a distance to the Virgo cluster of 16.5 Mpc. Dotted lines represent individual Gaussian fits while dashed lines are the sum of these fits.}
\label{lum-mue}
\end{figure}


The correlation of $\mu_{e,H}^i$ with rotational velocity gives
greater hope of establishing a link with the observed bimodality in
$\mu_e$. Indeed, we note in \Fig{vel-mue} hints of bimodality for
Virgo galaxies in the distribution of the maximum galaxy rotational
velocity, V$_{max}$. With high-quality, heterogeneous (combination of
\ion{H}{I} linewidths and H$\alpha$ rotation curves) dynamical
information available from HyperLEDA (Paturel et al. 2003) for only
102 of the 286 VCC galaxies, the observed weak correlation (r=0.53)
between V$_{max}$ and $\mu_{e,H}^i$ may not be surprising. A
homogeneous collection of resolved rotation curves for the complete sample
of galaxies will be needed to establish a dynamical extension of the
galaxy structural bimodality.  In a similar vein, TV97 found for a
subsample of 35 UMa galaxies that the surface brightness correlated
strongly with the slope of the rotation curve, a subject we address in
Appendix 2.  Yet again, a much larger sample is needed for a
conclusive analysis.

Finally, we observe the distribution of luminosities vs $\mu_{e,H}^i$
in \Fig{lum-mue}. There is a clear trend between luminosity and
surface brightness, as expected. Although there is scatter in both
variables, any bimodality in $\log V_{max}$, as observed in
\Fig{vel-mue} ought to be reflected in $\log L$ by virtue of the
velocity-luminosity relation of galaxies. Furthermore, since
$\mu_e\sim\left\langle \mu\right \rangle_e$, and $\left\langle
\mu\right \rangle_e=L_{1/2}/\pi r_e^2$ we would expect the $\mu_e$
bimodality to be reflected in the distribution of luminosities, unless
the SB bimodality is due to the relation $L/r^2$. \Fig{lum-mue} shows
a mild bimodality in L, but which is uncorrelated with the stronger
bimodality in $\mu_e$.

We have analyzed bivariate distributions involving color, rotation
velocity, luminosity and surface brightness. From
Figs. \ref{mue_gi}-\ref{lum-mue}, we find no strong correlation
between the structure seen in the distribution of surface brightnesses
and that of any other structural parameter. There is no correlation between surface brightness and
color, with both HSB and LSB galaxies occupying the red and blue
peaks. Figs \ref{vel-mue} \& \ref{vratio} suggest that the source of
the bimodality in surface brightness may be linked to galaxy dynamics
and not stellar populations. TV97 also proposed a dynamical explanation for the
structural bimodality. In the following section we offer a
tentative explanation for this exciting result in a more global
context.

\section{Discussion}\label{sec:discussion}

In \S4, we have verified that the distribution of surface brightnesses
is bimodal in UMa and Virgo cluster galaxies. Before we discuss possible explanations for this
effect, let us investigate any possible biases in the data measurements
and calibrations that could conspire to produce the observed bimodalities. 
Environment, which we discuss below, may also play a role in the 
structural bimodality.  Various angles to address any cause for bimodality 
- field vs cluster galaxies or optical vs IR data, or both - will then 
be explored.  We conclude with plausible explanations for the observed
galaxy surface brightness bimodality.

\begin{figure}
\centering
\includegraphics[width=0.48\textwidth]{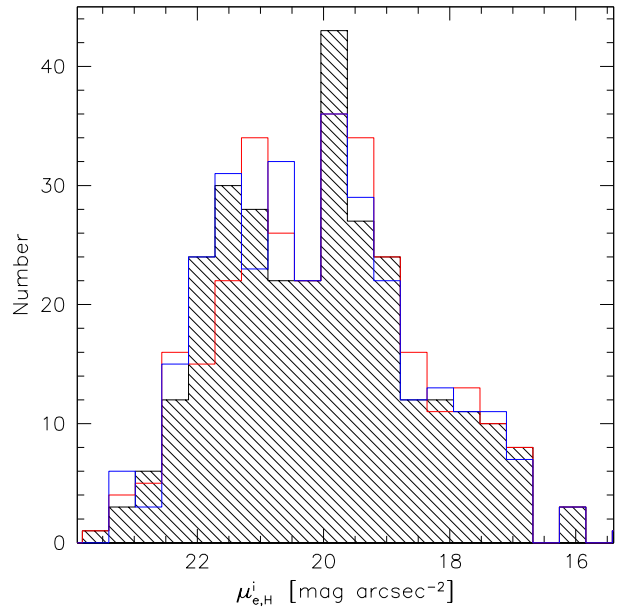}
\caption{Distribution of $\mu_{e,H}^i$ for 286 NIR VCC galaxies with
  an over-subtraction of the sky of 0.05\% (red), and
  under-subtraction of 0.05\% (blue) and the nominal sky subtraction
  (black).}
\label{esbskysub}
\end{figure}

\subsection{An Artificial Bimodality?}
We here investigate any potential source of bias from our measurement
and calibration techniques. In our data paper (McDonald et al. 2009),
we quote estimates for the error
induced by our photometric calibration of $\sim$0.1 H-$\magarc$.
In Figs. \ref{virgo_csb}a and \ref{virgo_esb_bymorph}, we observed 
that the separation between the LSB and HSB peaks is $\sim$2 H-mag
arcsec$^{-2}$. In order to artificially cause this gap, a surface
brightness error greater than 0.5 H-$\magarc$ on both LSB and
HSB peaks must be invoked. This error would also be systematic in
order to ``evacuate'' ISB galaxies: bright LSB galaxies would have to
be made fainter and/or faint HSB galaxies would have to be made
brighter. We will examine below three possible processes that could
induce such a large error: inclination correction, sky subtraction,
and surface brightness measurements. We will attempt to show that,
although far from negligible, these errors can be controlled and
should not bias the shape of surface brightness
distributions. Finally, we will examine the possibility that either of
the SB bimodalities are due to regions with clustered HSB or LSB
galaxies (i.e. an infalling group). 

\begin{figure}
\centering
\includegraphics[width=0.48\textwidth]{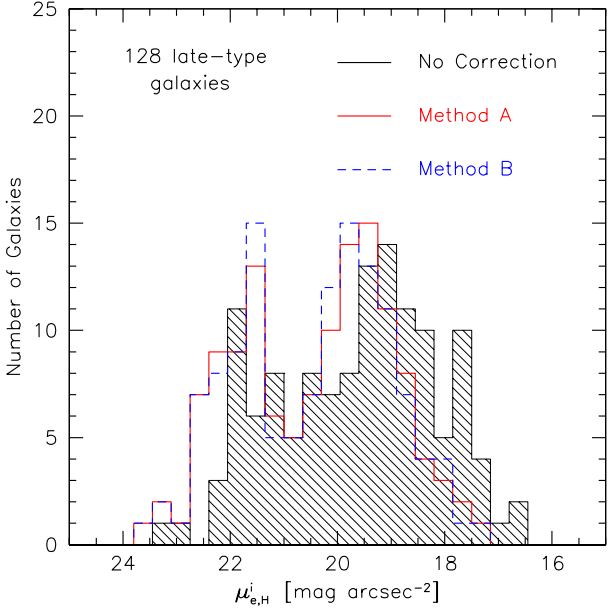}
\caption{Distribution of effective surface brightness for 128
  late-type VCC galaxies using different inclination corrections.}
\label{2incs}
\end{figure}

\subsubsection{Sky Subtraction Errors}
In Appendix A, we demonstrated how a sky measurement error of $\pm$0.01\% can 
result in an artificial truncation or upturn in a deep surface 
brightness profile. However, while LSB galaxies are highly sensitive
to sky errors, a systematic over- or under- subtraction of the sky could hardly 
enhance the ISB gap. In \Fig{esbskysub}, we show the distribution
of $\mu_e$ if the images for all the sample galaxies are 
systematically over- or under-subtracted by 0.05\% (a huge 
error at H-band). It is clear that, the net effect is to
shift all galaxies towards LSB, not to create a gap.

\subsubsection{Inclination Corrections}
In \S4.1, we discussed deprojection corrections to measured
surface brightness quantities. At NIR wavelengths, galaxies are nearly
transparent and the correction for deprojection is purely geometric. 

Since $\mu_0$ is measured from the disk fit, the geometric correction
is trivial, provided the inclination of the disk is known.  However, 
the quantity $\mu_e$, which depends on both bulge and disk light, is
less easily corrected. We thus make use of two different methods to 
correct $\mu_e$ for projection.  First, we can write the corrected 
effective surface brightness of the whole galaxy, $\mu_e^i$, as:
\begin{equation}
\mu_e^i=\mu_e - 2.5\log\left({b\over a}\right)_{r_e},
\end{equation}
where $\left(b/a\right)_{r_e}$ is the axial ratio at the effective
radius, $r_e$. This correction (Method A) is straightforward, since 
we know the ellipticity as a function of radius from isophotal fits
(\S3.2.3). 

The second inclination correction (Method B) requires the
decomposition of the SB profile into bulge and disk components and
deprojection of the two components separately. Knowing the inclination
of the disk, we can deproject the disk fit (assuming full transparency). 
We further assume that the bulge requires no deprojection. 
This yields the correction for the global
$\mu_e^i$:
\begin{equation}
\mu_e^i=-2.5\log\left\{I_{bulge}(r_e) + I_{disk}(r_e)\left({1\over{b/a}}\right)_{r_{max}}\right\},
\end{equation}
where $\left(b/a\right)_{r_{max}}$ is the axial ratio of the outermost
isophote. The effective radius is measured from the total light
profile and the contributions from the individual bulge and disk
profiles at this radius are used. This correction is intuitively
appealing as the bulge should require less of a deprojection
correction than the disk. The assumption of a spherical bulge may
however be inadequate, especially in the presence of a bar. However,
we find that the isophotes in the central regions of most of our
galaxies that exhibit bulges are, in fact, fairly round.

\Fig{2incs} shows the result of applying the two aforementioned
inclination corrections. We find very little difference
between methods A \& B. As expected, the corrected values to face-on
orientation for
$\mu_{e,H}^i$ are systematically fainter. Most importantly, regardless of the
correction used, bimodality is always preserved. As there is little
difference in the resulting distributions, we prefer Method A for its
simplicity and empirical, rather than parametric,
quality. Furthermore, Method B is less desirable since it may yield
large errors in cases when the disk fit is unreliable or if the
assumption of spherical bulge is unwarranted. 

\subsubsection{Surface Brightness Measurements}
\begin{figure*}
\centering
\includegraphics[width=0.9\textwidth]{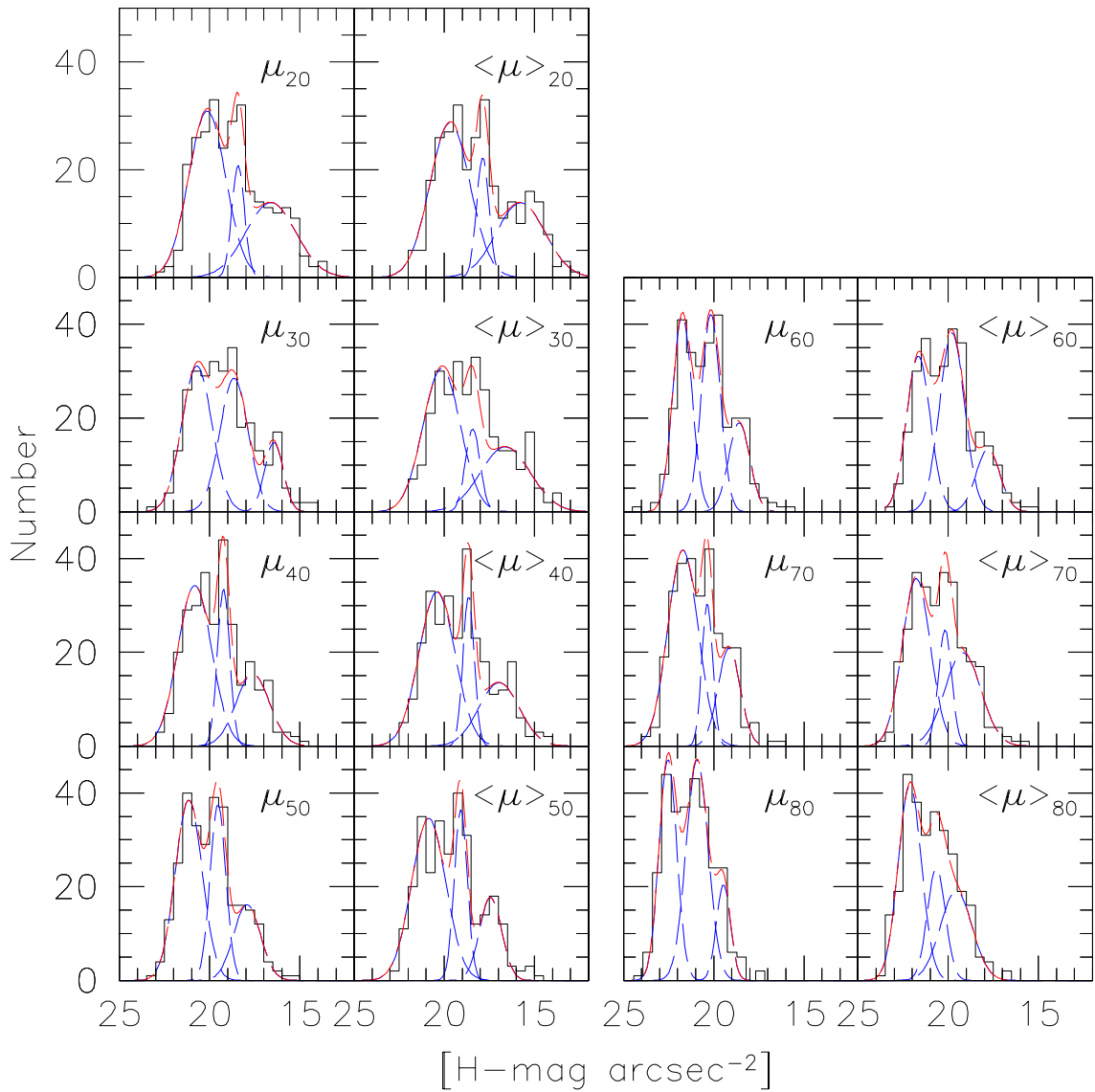}
\caption{Surface brightness distributions for local and global
  measurements at different galactocentric radii. $\mu_X$ refers to
  the surface brightness at the radius containing $X$\% of the total
  light, while $\langle \mu \rangle_X$ refers to the average surface
  brightness interior to that radius. The blue dashed lines are
  multiple Gaussian fits that correspond to the superimposed bimodal
  distributions of early- and late-type galaxies. The red solid line
  is the sum of the Gaussians.}
\label{allmues}
\end{figure*}

Ideally, we wish to avoid any sensitivity to the bulge light as well 
as to localized radial structure.  Such a parameter for pure disks 
is $\mu_0$, but we do not have an analog for both disks and spheroids.
Indeed, $\mu_e$ is affected by local structure while
$\left\langle \mu\right \rangle_e$ encompasses bulge light. Of interest to us is to
investigate if the use of a local vs global estimator of surface
brightness at different galactocentric 
radii should cause bimodality.  We can compute $\mu_X$ and
$\left\langle \mu\right \rangle_X$ as in \S3.2.2. At short radii, we expect both 
$\mu_X$ and $\langle \mu \rangle_X$ to be biased by any bulge or bar, while 
at large radii $\mu_X$ should suffer mostly from SB fluctuations
and sky subtraction errors. \Fig{allmues} and Table \ref{sbtable} 
show the results of our exercise for all 286 VCC galaxies in our
sample at H-band.  The result is interesting: regardless of the
specific measure of surface brightness, whether local or global,
computed at small or large radii, the three SB peaks are always preserved.

\begin{table}
\caption{Location and separations of the three SB peaks for SB
  measurements.}
\centering
\begin{tabular} {cccccc}
\hline\hline
SB Measured & ESB & HSB & LSB & E-H & H-L\\
\hline
$\mu_{20}$&16.61 & 18.42 & 20.14 &1.81 &1.72\\
$\mu_{30}$&16.44 & 18.65 & 20.71 &2.16 &2.06\\
$\mu_{40}$&17.61 & 19.23 & 20.83 &1.62 &1.60\\
$\mu_{50}$&17.93 & 19.54 & 21.16 &1.61 &1.62\\
$\mu_{60}$&18.59 & 20.16 & 21.74 &1.57 &1.58\\
$\mu_{70}$&19.15 & 20.36 & 21.76 &1.21 &1.41\\
$\mu_{80}$&19.46 & 20.92 & 22.54 &1.46 &1.62\\\\

$\left\langle \mu\right \rangle_{20}$& 15.76 & 17.86 & 19.68 &2.10 &1.82\\
$\left\langle \mu\right \rangle_{30}$& 16.61 & 18.42 & 20.14 &1.81 &1.72\\
$\left\langle \mu\right \rangle_{40}$& 16.96 & 18.65 & 20.40 &1.69 &1.75\\
$\left\langle \mu\right \rangle_{50}$& 17.43 & 19.08 & 20.86 &1.65 &1.78\\
$\left\langle \mu\right \rangle_{60}$& 17.80 & 19.79 & 21.66 &1.99 &1.87\\
$\left\langle \mu\right \rangle_{70}$& 19.19 & 20.18 & 21.77 &0.99 &1.59\\
$\left\langle \mu\right \rangle_{80}$& 19.56 & 20.66 & 22.13 &1.10 &1.47\\

\hline
\end{tabular}
\label{sbtable}
\end{table}


As can be seen from Table \ref{sbtable}, the separations between the
ESB and HSB peaks, and the HSB and LSB peaks are roughly equal. The
remarkable result that the peak separations $\langle $ESB-HSB$ \rangle$ and
$\langle $HSB-LSB$ \rangle$ are nearly identical at most SB levels suggests that
there may be a single mechanism responsible for all three peaks.

Thus, we have shown that both the early- and late-type structural bimodality
persist regardless of the surface brightness measurement, inclination
correction, and sky subtraction error.  Let us then address possible
causes for the observed galaxy bimodalities, convinced that they are
not an artifact of our methodology.

\subsubsection{Spatial Surface Brightness Distributions}
\begin{figure*}
\centering
\includegraphics[width=0.9\textwidth]{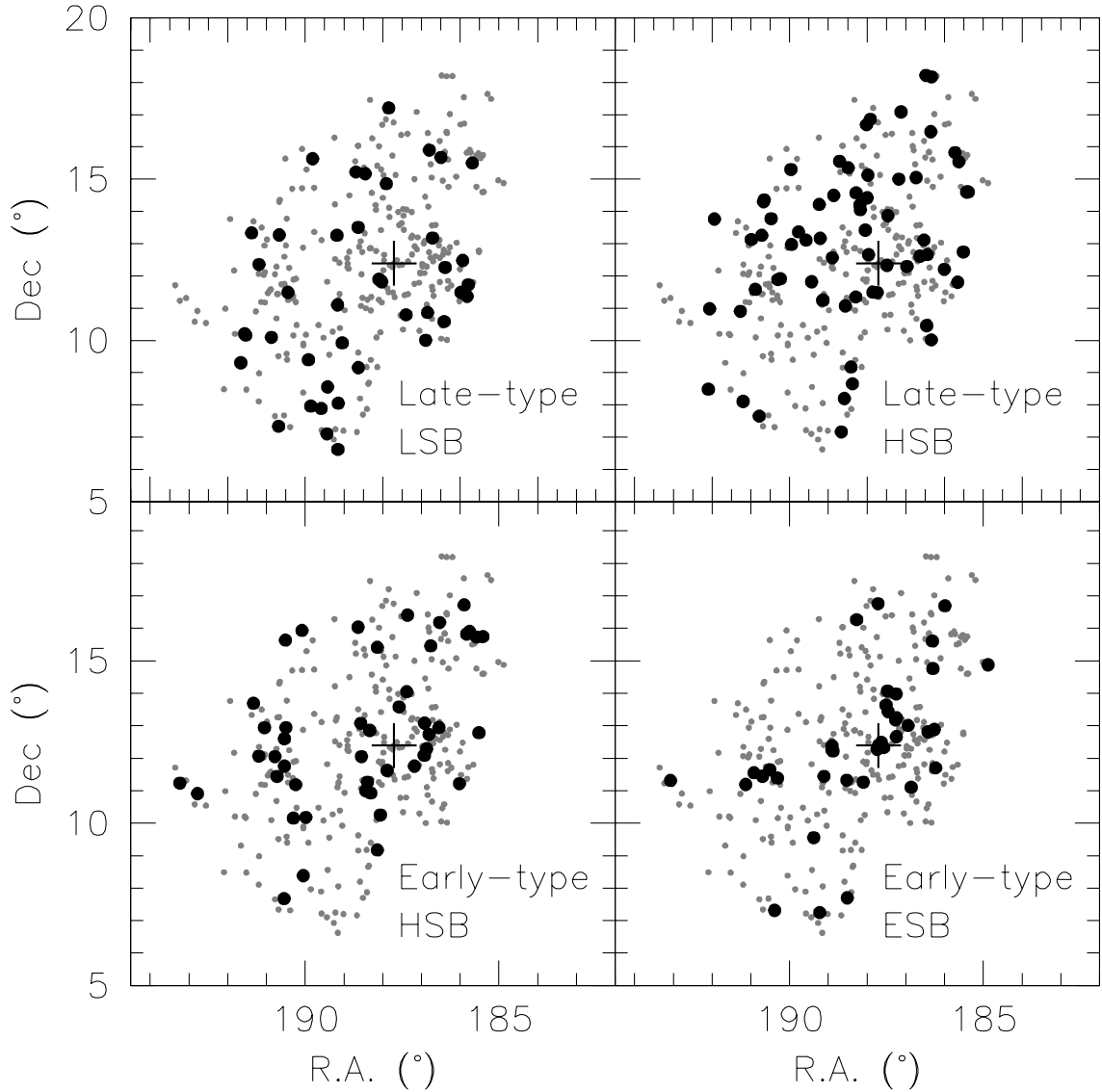}
\caption{Spatial distribution of VCC galaxies in four surface
  brightness bins: LSB late-type ($\mu_{e,H}^i>21$ mag arcsec$^{-2}$),
  HSB late-type ($\mu_{e,H}^i<21$ mag arcsec$^{-2}$), HSB early-type
  ($\mu_{e,H}^i>18.5$ mag arcsec$^{-2}$), and ESB early-type
  ($\mu_{e,H}^i<18.5$ mag arcsec$^{-2}$). The smaller gray points
  represent the full sample of 286 Virgo cluster galaxies.}
\label{esb_spatial}
\end{figure*}

It is conceivable that one of the three surface brightness
peaks could be due to the infall of a small group of galaxies
with a narrow surface brightness distribution. In order
to investigate this issue we turn to \Fig{esb_spatial}, which shows the spatial
distribution of galaxies in the four different SB peaks: HSB \& LSB early-type
galaxies, and HSB \& LSB late-type galaxies. It is clear from \Fig{esb_spatial}
that both the HSB and LSB spiral and irregular galaxies are
spread uniformly throughout the cluster. Likewise, most of the
spheroids are distributed uniformly throughout the cluster, with an
overabundance of HSB spheroids near the cluster center. Thus, although
there is a constant infall of galaxies into the Virgo cluster, it does
not appear that this is causing an artificial peak in the surface
brightness distribution.

\subsection{Cluster vs Field Distributions}
Freeman's law (1970) in its modern form is understood as the maximum
surface brightness any galaxy disk can have\footnote{Freeman's original
sample was heavily biased towards high surface brightness systems, hence 
the narrow reported range of surface brightnesses.}, but it does not 
preclude the existence of fainter systems. The general picture today 
is that the surface brightness function of galaxies has a tail at low
surface brightnesses that extends to arbitrarily low values 
(de Jong \& Lacey 2000). Previous studies that have attempted to quantify 
the distribution of galaxy surface brightnesses have focused mostly on
field systems (e.g., McGaugh \& de Blok 1995; de Jong \& Lacey 2000;
Driver \etal 2005). The effect of environment on galaxy surface 
brightnesses, if any, would thus be missed. 

In their pioneering study, TV97 observed that the structural bimodality 
is strengthened if their UMa sample is restricted to isolated galaxies.
For UMa, an isolated galaxy was defined as a galaxy that has not
encountered another galaxy during the lifetime of the cluster. Since
UMa is dynamically young, this definition is a reasonable
one. However, since the crossing time of the Virgo cluster is less
than one-tenth of a Hubble time (Trentham \& Tully 2002), it is
unreasonable to perform a similar analysis for Virgo
cluster galaxies. Every galaxy in the cluster should have had at least one significant
encounter within the last Gyr, which means that we cannot divide the
sample into isolated and non-isolated galaxies.

Various studies have reported that LSB galaxies form primarily in
isolated environments (e.g., Rosenbaum \& Bomans 2004) likely from
1-2$\sigma$ peaks in the initial density fluctuation spectrum. 
However, these authors found that local LSBs are almost exclusively 
found in clusters and filaments.  
If LSB galaxies form from 1-2$\sigma$ peaks, they would migrate 
from voids to high-density regions. This does not explain how 
dynamically young galaxy clusters, such as UMa, could have a high abundance
of LSB galaxies. Instead, perhaps the cluster environment enhances
LSB formation which would explain the current location of the LSB 
population as well as the SB bimodality seen in UMa and Virgo.

For the UMa and Virgo clusters, there is a bimodality of 
disk surface brightness in spiral galaxies as well as bimodality 
in effective surface brightness for early-type galaxies.  Since 
previous studies of field galaxies have not shown any evidence 
for bimodal surface brightness distributions, we might therefore 
think of structural bimodality as a cluster effect. However, it 
should be noted that most of the studies of the surface brightness distribution
of field galaxies have focused strictly on optical imaging for lack
of available complete similar data bases at IR wavelengths. We showed
in Figs. \ref{virgo_sdss_csb} \& \ref{virgo_esb_sdss} that the
bimodality in disk and total surface brightnesses is enhanced in
the NIR, due to uncertain and significant corrections for dust at
optical wavelengths. None of the major NIR data bases available 
today (e.g., de Jong 1996; Jarrett \etal 2003; Grauer \etal 2003; 
MacArthur \etal 2003; Smith \etal 2008) cover the full range of surface brightness to sample the LSB peak. 
For this reason, it is unclear whether the structural bimodality 
observed at NIR wavelengths is indeed a cluster effect since it
has only been observed in galaxy clusters thus far, or whether it is 
inherent to galaxies in all environments. Since we see structural 
bimodality in the most isolated galaxies of the UMa cluster as 
well as in the denser environment of Virgo, it appears that this 
effect is independent of environment. New data from field NIR 
surveys such as PanSTARRS and UKIDSS (e.g., Smith \etal 2008) 
will be needed to revisit this question for field galaxies 
(provided careful assessment of completeness for field samples).

We now entertain possible physical interpretations for structural
bimodality.

\subsection{Differences Between Cluster and Field Galaxies}\label{sec:environment}

Dressler (1980) first demonstrated the existence of strong
correlations between various galaxy properties and environment. Red
sequence galaxies live preferentially in dense environments, whilst
blue, star-forming galaxies are predominantly found in isolation. This
dependence on environment was reinforced as a major
cause for the observed SDSS color and star formation bimodality. Not
surprisingly, the local density of galaxies causes a strong
morphological bias. Merging and tidal interactions, which are more
common in dense environments (Moss \& Whittle 1993; Gnedin 2003),
can deplete the gas content of spiral and irregular galaxies, turning
them into ellipticals and possibly S0s (Toomre \& Toomre 1972; 
Farouki \& Shapiro 1981; Barnes 1999). An intracluster 
gas can also strip gas from late-type galaxies, reducing star 
formation drastically (Gunn \& Gott 1972). Processes such as galaxy-galaxy harassment, tidal stripping, and starvation, which are more efficient
in dense environments, will all reduce the amount of active star formation. 
It is then not surprising that there should be a correlation between 
color, star formation and environment in this context. 

There are also several properties for the dark halo that appear to be
dependent on environment. The mass, shape, spin and formation time of
dark matter halos formed in denser environments that become clusters
or filaments appear to be significantly different than those formed in
isolation (Hahn \etal 2007). Simulations suggest that galaxy halos
with masses M $\lesssim$ M$_*$, where M$_*$ is the galaxy halo mass at
which the mass function turns over, have parameters that correlate
strongly with environment, while those with M $\gtrsim$ M$_*$ do
not. Hahn \etal (2007) find that low mass halos are older and have
higher angular momenta when formed in proto-cluster environments,
while low mass halos in the field form later and have lower angular
momenta.  The dependence of halo properties on environment could
explain part of the observed differences in cluster and field
galaxies.

However, using a sample of 329 nearby cluster and field star forming 
galaxies, Vogt \etal (2004) failed to find any correlation between 
stellar mass-to-light ratios or circular velocities with environment. 
More observational and theoretical studies of environmental effects 
on galaxies will be needed. 

\subsection{Theoretical Interpretation}
We have found so far that the gas-poor and the gas-rich UMa and Virgo 
cluster galaxies each exhibit a surface brightness bimodality. It is
unclear whether this result is unique to clusters, since there is no
complimentary NIR survey of field galaxies that reaches the same
depth in surface brightness. However, UMa very much resembles the
field due to its low density (see TV97), whilst Virgo is a much richer
cluster, suggesting that environment may not play a strong role.

This surprising result requires an
explanation which current models of galaxy formation fail to
provide. However, our results may suggest routes for further
investigation and allow the rejection of certain hypotheses. We now
address various explanations for the structural bimodality in light of
the available data and current theoretical ideas.

\begin{enumerate}
\item \textit{Stellar Populations}:
Previous studies based on SDSS imaging for very large number of
galaxies have shown bimodal distributions of color and star formation
rate (e.g., Strateva \etal 2001), which is largely the result of two
galaxy populations: blue, star-forming spiral galaxies and red,
quiescent spheroidal galaxies. We showed in \S4.5 that color is
uncorrelated with the observed surface brightness bimodalities of
cluster galaxies; stellar populations are thus unlikely to be
a factor.

\item \textit{LSB Formation Enhanced in Clusters}: 
If the structural bimodality is only present in clusters, it could
result from more efficient tidal torquing in dense environments (Barnes
\& Efstathiou 1987; Bett \etal 2007; Hahn \etal  2007). If dark matter
halos acquire more angular momentum in a proto-cluster environment, then LSB galaxies
(which halt their collapse early due to high angular momentum) would
preferentially be formed in these dense environments. In this
scenario, higher surface brightness galaxies could form from the
merging of lower SB galaxies if they contain gas that can dissipate 
energy before forming stars. One would thus expect a ``smearing''
of the LSB peak towards HSB, but a discrete gap is still unaccounted
for. It has been well documented that there is a maximum
surface brightness that galaxy disks will not exceed (Freeman
1970). If this maximal disk surface density is obeyed, then galaxies
which evolve at various rates from low to high surface brightnesses
via merging may ultimately halt their evolution at a disk surface density
equal to the Freeman value, thus enhancing the HSB peak. This scenario
does not preclude the formation of ISB galaxies via the merging of two
LSB galaxies or other such events, but simply suggests that the ISB
case might be a rapid transition stage for galaxies as they merge and evolve
towards the Freeman limit. Indeed, we do not see an absence of ISB
galaxies in clusters, but rather a relative excess of HSB and LSB galaxies. 

Our group is currently studying the velocity function (and
thus the distribution of angular momenta) of Virgo cluster galaxies
which will provide further insight into this problem. More studies like
that of Hahn \etal (2007) would help elucidate the role of environment on the
distribution of various dynamical properties.

\item \textit{Specific Stable Radial Configurations}: 
Expanding upon the ideas of Mestel (1963),
TV97 suggested that galaxy disks could settle into one of two stable
radial configurations: disks with a dynamically important baryonic
component in the center and those without (also known as maximal and
minimal disks). They suggested that intermediate surface brightness
galaxies are transient phenomena in high-density environments where interactions
re-arrange the baryons. This is supported by the fact that, in UMa,
most intermediate surface brightness galaxies have significant
neighbors and thus the bimodality is enhanced for isolated
systems. However, the lack of ISB galaxies in the densest
regions of the Virgo cluster suggests the opposite result.

If the structural bimodality is unaffected by environment then there
may be a more fundamental mechanism preventing the stability of ISB
galaxies. The ISB regime may correspond to an unstable dynamical
configuration, possibly implying a bimodality in the spin parameter,
$\lambda$, for disk galaxies. 

\item \textit{Gas Depletion in Clusters Halts Evolution}: 
The amount of gas retained by cluster and field galaxies differ due to
various factors such as high frequency of tidal interactions in
clusters (Moss \& Whittle 1993; Gnedin 2003) and the dense
intracluster medium which can cause gas stripping (Gunn \& Gott
1972). If ISB galaxies form primarily via passive evolution and HSB
galaxies form primarily via mergers, then the cluster environment
would act to prevent LSB-to-ISB evolution and enhance LSB-to-HSB
evolution. 

\end{enumerate}

Without being able to pinpoint the role that environment plays in the
observed structural bimodalities, it is difficult to provide an
explanation for their existence. It is likely that no single
hypothesis above fully explains the surface brightness bimodality and
that the separate bimodalities in early-type and late-type galaxies
will require their own explanations (although their alignment and
equal spacing, as seen in Table \ref{sbtable}, suggests something
fundamentally similar!). It is difficult to speculate further without
a proper analysis of the distribution of $\mu_e$ for field galaxies, a
complimentary dynamical analysis of the Virgo cluster, as well as
detailed numerical simulations for the effects of external processes,
such as tidal stripping and mergers, or internal processes, such as
outflows, on surface brightness profiles of cluster and field
galaxies.

\section{Conclusion}
We have constructed the deepest complete sample of near-IR photometry
of Virgo cluster galaxies to date and with it, we were able to achieve
a detailed analysis of the distribution of surface brightness profiles
and luminosities for 286 Virgo cluster galaxies. We also extracted $griz$
photometry from the SDSS for comparative analysis with a larger sample
of 742 VCC galaxies. This Virgo cluster study also follows from our
re-analysis of similar data for the smaller UMa cluster (McDonald
\etal 2008). Our main findings are:
 
\begin{itemize}
\item There exists a structural bimodality in each of the early and
  late type galaxy populations of the UMa and Virgo clusters. From NIR
  images of these galaxies, we observe surface brightness peaks at
  $\mu_{e,H}^i$=21.7 $\magarc$ and $\mu_{e,H}^i$=19.7 $\magarc$ for
  late-type galaxies, and $\mu_{e,H}^i$=19.7 $\magarc$ and
  $\mu_{e,H}^i$=17.8 mag arcsec$^{-2}$ for early-type galaxies. These
  three peaks are independent at a confidence level $>$90\%.

\item The observed structural bimodalities in each galaxy class (early
  or late) is independent of the observed color bimodality. 

\item The NIR luminosity function for Virgo has a prominent dip at
  intermediate luminosities which is also seen, to a lesser extent, in
  the optical SDSS data.


\end{itemize}

The fact that the dichotomy between HSB and LSB galaxies is
independent of color and morphology implies a possible dynamical
connection. We have detected a weak, but non-negligible, correlation
between $V_{max}$ and $\mu_e$, but more data will be needed to
strengthen any dependence.  
The range of densities from UMa to Virgo
also suggests that the mechanism for producing HSB and LSB galaxies is
independent of environment. Without a sample of true field galaxies
it would, however, be unwise to speculate about any environmental
effect on $\mu_e$. Unfortunately, current studies of the
distribution of surface brightnesses of field galaxies are either
hampered by dust extinction in the optical or have not achieved deep 
and complete enough coverage at NIR wavelengths. Thus, most reasonable
scenarios for field/cluster galaxy evolution are still plausible. 

The most obvious and relevant test for the hypothesis of different
$\lambda$-distributions in cluster and field environments, as proposed
by Hahn \etal (2007), is to
assemble a complete sample of resolved rotation curves for cluster
galaxies. This would enable new constraints on the distribution of
the spin parameter, $\lambda$, which may be responsible for
setting $\mu_e$ (e.g., Dutton \etal 2007). Hahn \etal (2007) predict a 
distribution that is skewed towards high values of $\lambda$ in clusters
and filaments. A complete sample of resolved dynamical data for
cluster and field galaxies should allow a full assessment of the
observed structural bimodality in terms of unbiased parameters such as
dynamical mass and the slope of the rotation curve.

It has been over ten years since the discovery of the surface brightness
dichotomy in disk galaxies by TV97. We have now extended that study 
to a broader environment to establish the presence
of three separate peaks in the distribution of effective surface
brightnesses in all galaxies with $>$90\% confidence, and two peaks in
the distribution of disk central surface brightnesses in disk galaxies
with 95\% confidence. The next few years should see significant
improvements in the size and scope (e.g., range of surface brightness)
of galaxy NIR surveys and in studies into the effect of environment on
$\mu_e$. The theoretical interpretation for these results however
awaits a better determination of $\lambda$ for galaxies in the field
and in clusters, and a better interpretation of the combined effects
of merging and stripping to the overall surface brightness of galaxies.

\section{Acknowledgements} 
We acknowledge valuable conversations with Jeremy Bailin, Aaron
Dutton, Julianne Dalcanton, Roelof de Jong, Lauren MacArthur,
Hans-Walter Rix, Frank van den Bosch, and Sylvain Veilleux.  Many
thanks also to Guiseppe Gavazzi for providing precious GOLDMine galaxy
images, to Yucong Zhu for providing SDSS luminosities for the VCC/SDSS
sample, and to Joel Roediger for his stellar mass-to-light ratios for
the Virgo ``H-band'' galaxies. We are also grateful to Jon Loveday for
his careful referee report which led to an improved presentation of
our paper.

SC would like to acknowledge financial support 
via a Discovery Grant from the National Science and Engineering Council  
of Canada. RBT acknowledges support from US National Science
Foundation award AST 03-07706.
 
This research has made use of $(i)$ the NASA/IPAC Extragalactic Database  
(NED) which is operated by the Jet Propulsion Laboratory, California  
Institute of Technology, under contract with the National Aeronautics  
and Space Administration, as well as NASA's Astrophysics Data System;  
$(ii)$ the $Sloan$ $Digital$ $Sky$ $Survey$ (SDSS). Funding for the  
creation and distribution of the SDSS Archive has been provided by  
the Alfred P. Sloan Foundation, the Participating Institutions, the  
National Aeronautics and Space Administration, the National Science  
Foundation, the U.S. Department of Energy, the Japanese Monbukagakusho,  
and the Max Planck Society. The SDSS Web site is http://www.sdss.org/.  
The SDSS is managed by the Astrophysical Research Consortium (ARC) for  
the Participating Institutions.  
$(iii)$ the $HyperLeda$ database (http://leda.univ-lyon1.fr).


\newpage 

\clearpage 

\clearpage 

\clearpage 

\clearpage 

\clearpage 

\clearpage 


\clearpage 
\clearpage 

\clearpage 

\clearpage 

\clearpage 

\clearpage 

\clearpage 

\clearpage

\clearpage 




\newpage 

\appendix 

\section{A. Model Expectations}\label{sec:models}

\begin{figure*}
\centering
\includegraphics[width=0.9\textwidth]{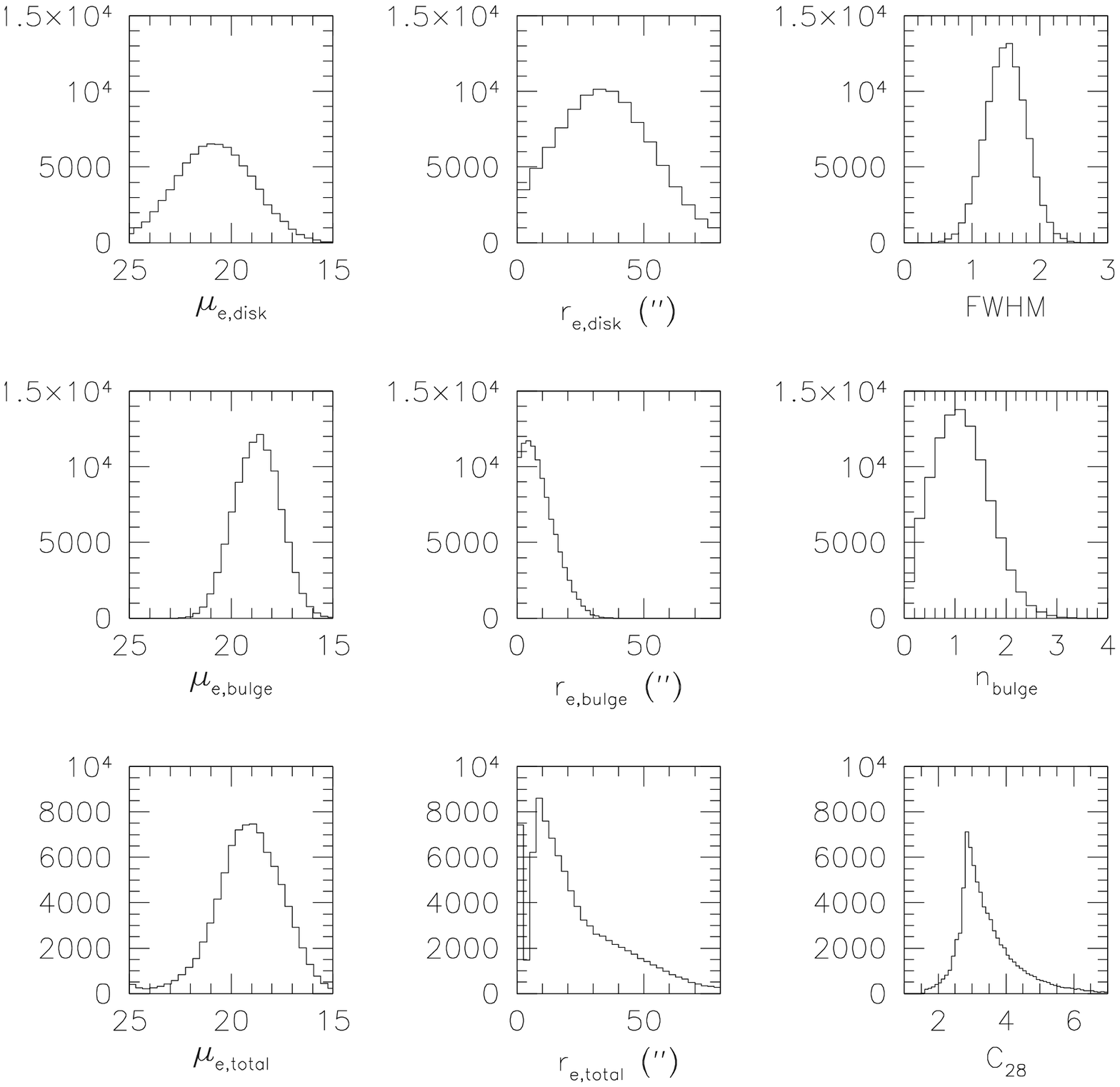} 
\caption{Distribution of input parameters (top 6 panels) and measured
  parameters (bottom 3 panels) for 100,000 model surface brightness
  profiles.}
\label{model-params}
\end{figure*}

In order to characterize the intrinsic distribution of structural 
parameters such as $\mu_e$, $r_e$ and $C_{28}$, free from evolutionary 
effects, we have generated a Monte Carlo suite of 
idealized galaxy models using combinations of a large number of model 
parameters.  A single galaxy light profile can be determined
with 8 basic parameters. For the bulge and disk, the parameters 
$\mu_{e,d}$, r$_{e,d}$, $\mu_{e,b}$, r$_{e,b}$ and \Sersic $n$ are 
selected randomly from Gaussian distributions whose widths 
match roughly observed galaxy distributions.  Another 2 parameters 
govern the location and brightness of spiral arms.  The location 
of the arms varies from 1 to 2 disk scale lengths and the peak arm 
brightnesses ranges from 0.3 to 0.5 $\magarc$ above the disk brightness. 
Finally, all model profiles were convolved with a Gaussian function to 
simulate the effect of atmospheric blur; the seeing FWHM is our
8th model parameter.  We have also forced 1 in every 5 profiles to 
be a pure-Sersic function to reflect the presence of systems with
no disk. The Gaussian distribution of input parameters for our 
models is shown in the top 6 panels of \Fig{model-params}.

In order to infuse more realism into this experiment, two different
types of error were added to the model profiles: photon (galaxy + sky)
random noise and systematic sky subtraction error.  We determined the
appropriate photon noise in the H-band from the distribution of
measured surface brightness errors for our H-band images 
(McDonald \etal 2009).  From this distribution, an error function 
was determined to add cosmic error to our model surface brightness 
profiles:

\begin{equation}
\Delta(\mu)=0.00075\times\exp{\left({\mu-16.5}\over{1.2}\right)} + 0.014~\magarc.
\end{equation}

This result is very similar to that produced by Courteau (1996) for
$r$-band surface brightness profiles. A typical noise value $\pm\Delta(\mu)$
now is added randomly to each model profile at each surface 
brightness level. 

In order to simulate sky-subtraction errors, a systematic ``measurement
error'' was first determined. This value is $\pm0.003\%$ of the sky
brightness at H-band.
This random sky error can artificially produce truncated 
(over-subtracted sky) or upbending (under-subtracted sky) 
surface brightness profiles, depending on its sign.

We have generated model galaxies by combining the 8 model parameters, each
selected randomly from a realistic range of values, and accounting for
random and systematic errors. The recovered parameter distribution for the
100,000 simulated surface brightness profiles can be seen in the lower
3 panels of \Fig{model-params} for the total galaxy light.  The
$\mu_e$ distribution is nearly Gaussian with a peak slightly brighter
than the disk $\mu_e$ peak due to the inclusion of substantial bulges
in most profiles. An HSB tail can be seen in the $\mu_e$ distribution
due to severely bulge-dominated systems. These systems have
higher $\mu_e$ values since bulges typically have higher surface
brightnesses than disks. The distribution of the $r_e$ parameter 
is a combination of bulge and disk components; the peak is 
controlled by the bulge $r_e$ for bulge-dominated systems and 
the tail is due to disk-dominated systems. Finally, there is a 
strong peak in the $C_{28}$ distribution near $C_{28}$=2.8, the expectation
value for pure-disk systems. Profiles with $C_{28}$$<2.8$ arise from 
cored ($n<1.0$) systems or galaxies with spiral arms.  
Profiles with $C_{28}$$>2.8$ or higher are bulge-dominated. $C_{28}$
values cover the full range from 1.5 to 6.5, in excellent agreement
with our data (see \se{conc}).

\begin{figure*}
\centering
\includegraphics[width=0.9\textwidth]{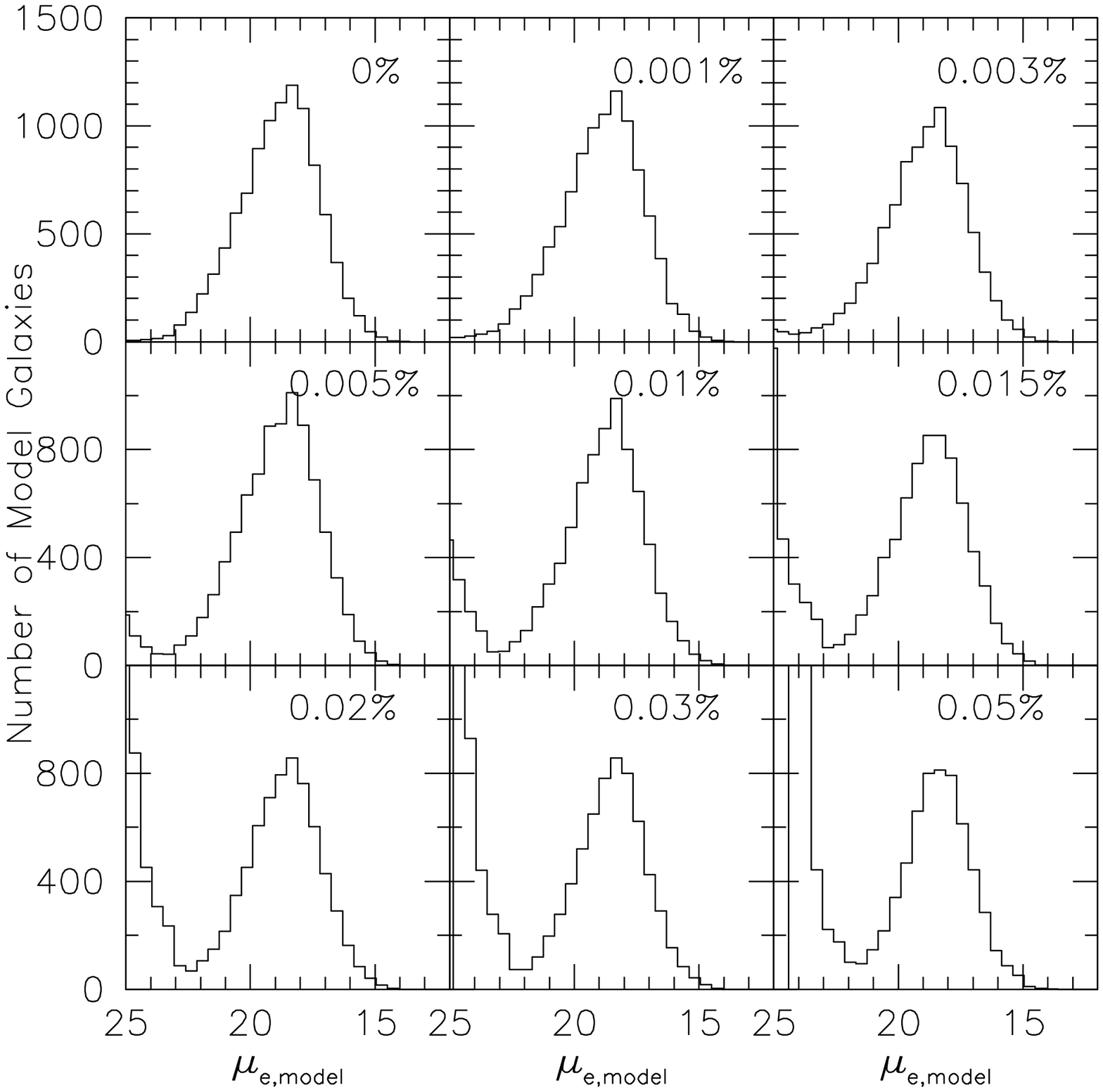}
\caption{Histogram of effective surface brightnesses of
  model profiles. The fraction in the upper right is the
  simulated sky error.}
\label{skyerr}
\end{figure*}

With our realistic model surface brightness profiles, we can trace the
effects of sky measurement errors. This is important since such errors
can drastically affect the shape of the density profile, causing an artificial
truncation or anti-truncation at large radii. The distribution of
total $\mu_e$ could thus be biased as a result. In \Fig{skyerr},
we show the effect of increasing the sky measurement error from
$\pm$0.001\% to $\pm$0.05\%. The net effect of a positive sky error 
is to pull LSB galaxies below the sky noise yielding a faint tail,
but the bright SB peak never shifts. 

\begin{figure*}
\centering
\includegraphics[width=0.9\textwidth]{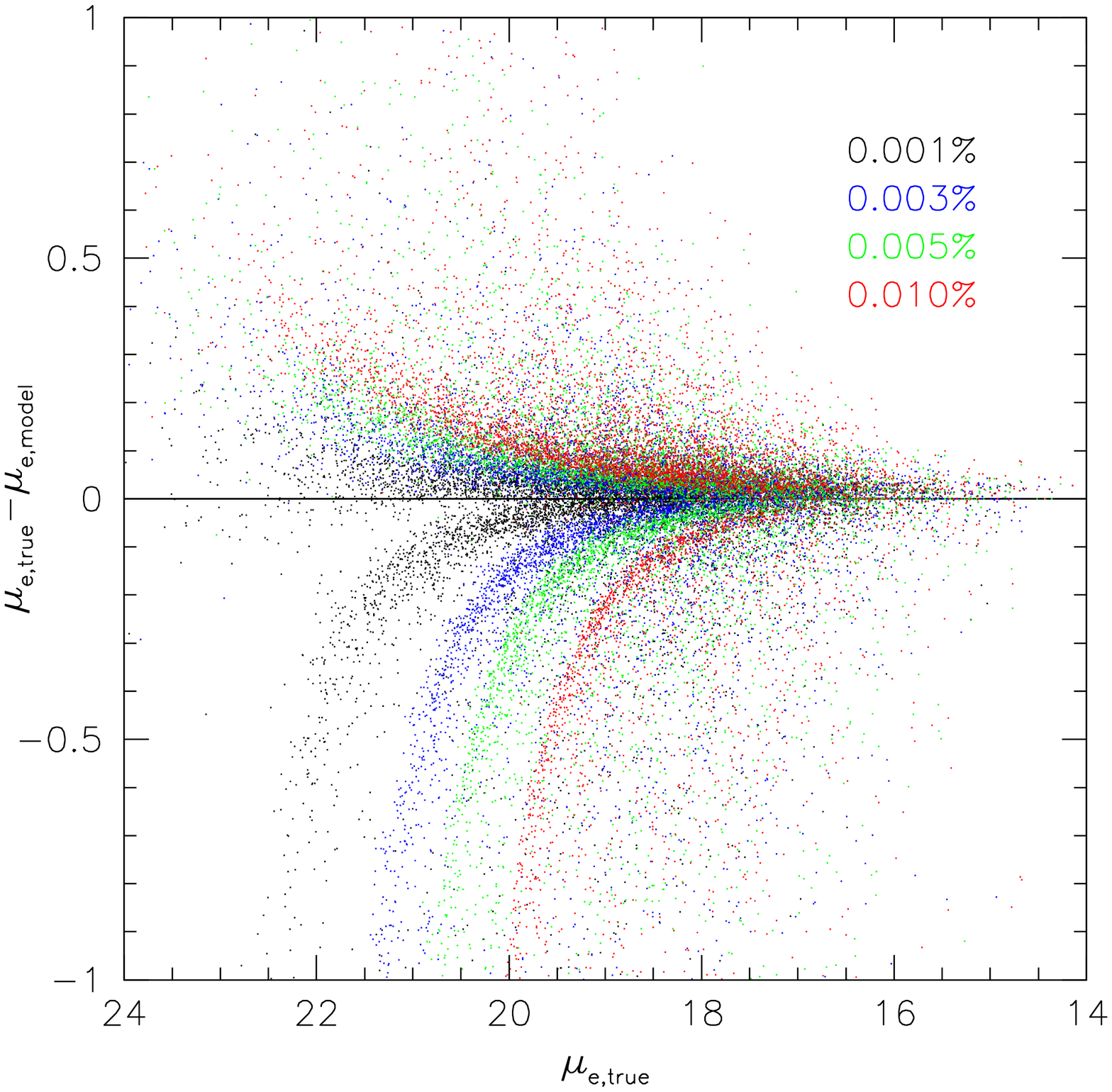}
\caption{Scatter on $\mu_e$ introduced by various sky estimate errors.}
\label{skymue}
\end{figure*}

\Fig{skymue} shows the variations in $\mu_e$ due to over- and 
under-subtracted sky errors. An under-subtraction error (upper half) 
rarely affects the measured $\mu_e$ by more than 0.2 $\magarc$. An 
over-subtraction error (lower half) has a more drastic effect 
as it can shift a significant part of the outer disk below 
observable levels. Recall that, while these tests about sky 
errors can be illuminating, the measured H-band sky error is
typically $\pm$0.005\% and the effect of a systematic sky error is thus 
minimal. 

Our simulations confirm that if the input structural parameters 
are distributed normally, we should expect a nearly normal 
distribution of effective surface brightnesses. Under no conditions
can we generate a distribution of $\mu_e$ that deviates
significantly from the normal function.  This result can be used 
as an unbiased comparison against galaxy structural parameters 
extracted from real data.

\clearpage 

\clearpage 

\clearpage 

\newpage
\section{B. Galaxy Dynamics}\label{sec:AppendixB}
The study of the distribution of galaxy structural parameters from
light profiles can benefit greatly from the availability of dynamical
information. For instance, TV97 found a clear relationship between a
galaxy's disk central surface brightness and the inner slope of its
rotation curve. In order to test this claim, and other correlations
between dynamical and photometric parameters, resolved rotation curves
would be ideal.  Failing any information about the shape of the
rotation curves, we resort to 1D measures of maximum
rotational velocity. Provided sufficient accuracy, they can, at
least, enable global comparisons with light profile parameters
for the spiral galaxies in our sample. 

We retrieved \hi data for 58 VCC galaxies from the Extragalactic
Distance Database (EDD; http://edd.ifa.hawaii.edu/) which overlap with
our H-band sample. These data have been corrected for aperture and 
turbulent effects following Tully \& Fouqu\'{e} (1985) who defined the linewidth parameter $W_R$.  
We have also verified that the linear (almost 1-to-1) relation between
radio and optically measured line widths observed by Courteau (1997) 
and others is also observed in this sample.   All linewidths 
were deprojected as $V_{tot}=0.5W_R^i=0.5W_R\sin{i}$ and $\sin i$ is given by: 

\begin{equation}
\sin^2{i} = {1 - q^2\over{1 - q_0^2}}.
\end{equation}
           
Here $q$ is the galaxy's axial ratio and $q_0=0.2$ is the estimated instrinsic thickness of the disk.

In order to duplicate Fig. 14 from TV97, we are interested in
computing the ratio $V_*$/$V_{tot}$. Here $V_*$ is the
maximum rotation velocity resulting from a purely baryonic system, 
while $V_{tot}$ is the observed rotation speed (which includes dark
matter) at a fiducial radius. TV97 postulate
that LSB galaxies tend to be dominated by dark matter all the way into
the innermost regions ($V_* \ll V_{tot}$) while HSB galaxies are
baryon-dominated within roughly two disk scale lengths ($V_* \sim
V_{tot}$; e.g., Courteau \& Rix 1997; D07).

To compute V$_*$, we use the following equation:
\begin{equation}
V_{*}(R)=\sqrt{{{GL(<R)}\over{R}}\times M_*/L_H}
\end{equation}

where $L(<R)$ refers to the luminosity within the radius, $R$, and
$M_*/L_H$ is the stellar mass-to-light ratio computed using SDSS
colors and Bell \etal (2003). The peak value of V$_{*}(R)$ is the
value we call V$_*$.

\begin{figure}
\centering
\includegraphics[width=0.48\textwidth,angle=0]{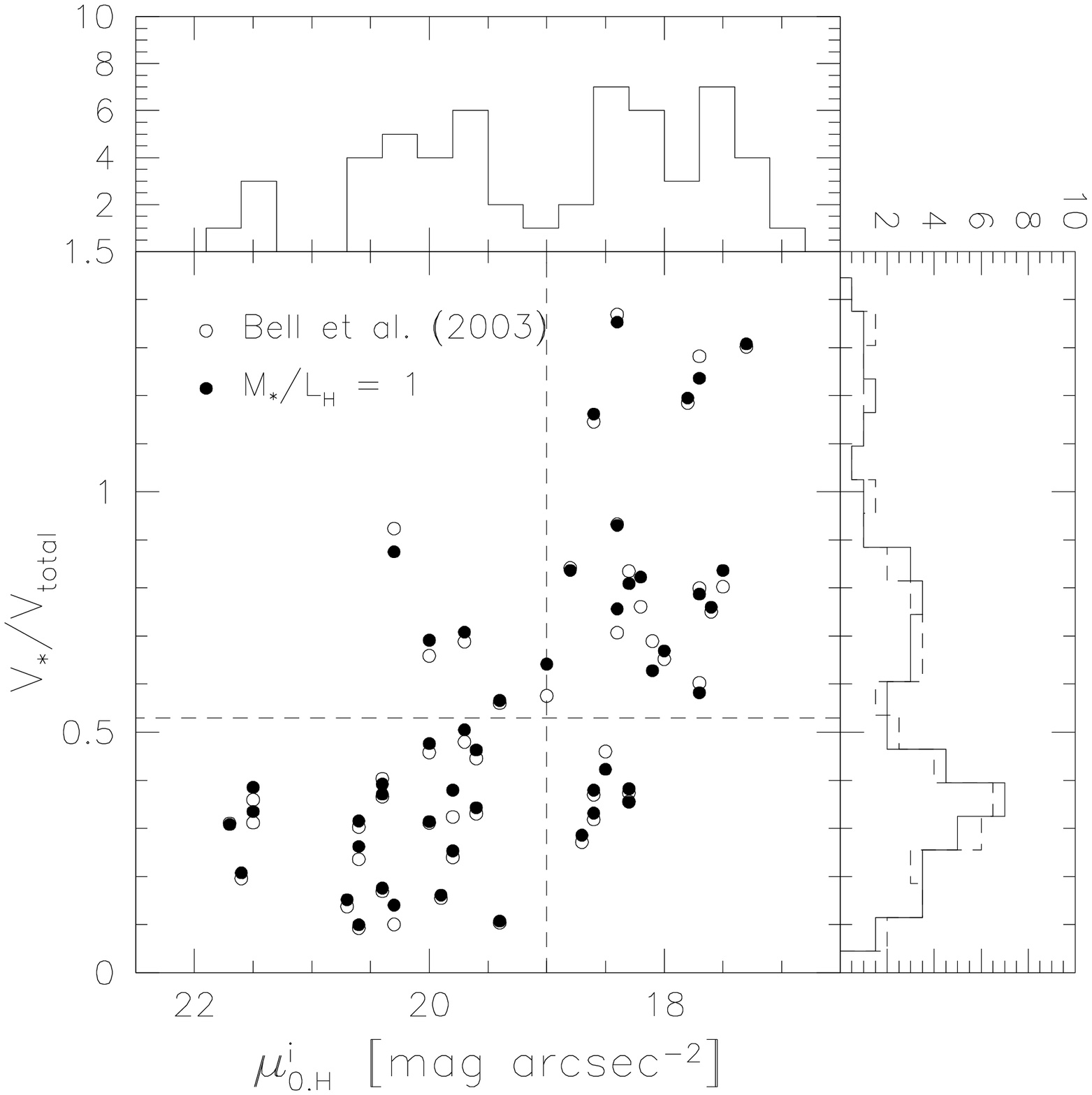}
\caption{Bivariate distribution of V$_*$/V$_{tot}$ and $\mu_{e,H}^i$
  for 58 disk galaxies in the Virgo cluster for two different
  mass-to-light ratios. The dashed histogram corresponds to the Bell et al. (2003) mass-to-light ratios, while the solid histogram corresponds to a constant mass-to-light ratio of 1.0.}
\label{vratio}.
\end{figure}

\Fig{vratio} shows the bivariate distribution of V$_*$/V$_{tot}$ as a
function of disk central surface brightness, $\mu_{0,H}^i$ for the spiral
and irregular galaxies with available linewidths from the EDD. 

We do not see the strong dynamical bimodality that is apparent in
TV97, however there does appear to be a correlation between
V$_*$/V$_{tot}$ and $\mu_{0,H}^i$, as well as hints of a bimodality in
$\mu_{0,H}^i$. LSB galaxies tend to be dark matter-dominated
(V$_*$/V$_{tot}<0.5$), while HSB galaxies would be baryon-dominated
(V$_*$/V$_{tot}>0.5$). The galaxies with V$_*$/V$_{tot}>1$ tend to
have bright bulges or compact nucleii which cause a spike in the
light-weighted rotation curve at short radii. However, with linewidths
for only 20\% of our sample, it is impossible to make these claims
with statistical significance. To do
this analysis correctly, we need to either obtain reliable linewidths
or deep, resolved, optical rotation curves for the full sample. Our
team is currently pursuing a complete characterization of the velocity
function of Virgo cluster galaxies in order to address these issues.

\label{lastpage}
\end{document}